\documentclass[12pt]{article}
\usepackage[a4paper,margin=1in]{geometry}
\usepackage[titletoc,title]{appendix}
\usepackage{changepage}
\usepackage[utf8x]{inputenc}
\usepackage{textcomp,marvosym}
\usepackage{fixltx2e}
\usepackage{amsmath,amssymb}
\usepackage{cite}
\usepackage{nameref,hyperref}
\usepackage{authblk}
\usepackage{algorithm}
\usepackage{algpseudocode}
\usepackage{dsfont}
\usepackage{xcolor}
\usepackage{enumitem}
\usepackage{caption,setspace}
\usepackage{subcaption}
\usepackage{pdflscape}
\usepackage{graphicx}
\usepackage{placeins}
\usepackage{mathtools}

\DeclareMathOperator\arctanh{arctanh}

\renewcommand{\eqref}[1]{Equation~(\ref{#1})}

\def\cbl{\color{black}}
\def\cb{\color{black}}

\title{Invading and receding sharp--fronted travelling waves}

\author[1]{Maud El-Hachem}
\author[1]{Scott~W. McCue}
\author[1]{Matthew~J. Simpson\footnote{To whom correspondence should be addressed. E-mail: matthew.simpson@qut.edu.au}}

\affil[1]{School of Mathematical Sciences, Queensland University of Technology, Brisbane, Queensland 4001, Australia}

\begin{document}

\maketitle
\begin{abstract}
Biological invasion, whereby populations of motile and proliferative individuals lead to moving fronts that invade vacant regions, are routinely studied using partial differential equation (PDE) models based upon the classical Fisher--KPP equation.  While the Fisher--KPP model and extensions have been successfully used to model a range of invasive phenomena, including ecological and cellular invasion, an often--overlooked limitation of the Fisher--KPP model is that it cannot be used to model biological recession where the spatial extent of the population decreases with time.  In this work we study the \textit{Fisher--Stefan} model, which is a generalisation of the Fisher--KPP model obtained by reformulating the Fisher--KPP model as a moving boundary problem.  The nondimensional Fisher--Stefan model involves just one parameter, $\kappa$, which relates the shape of the density front at the moving boundary to the speed of the associated travelling wave, $c$.  Using numerical simulation, phase plane and perturbation analysis, we construct approximate solutions of the Fisher--Stefan model for both slowly invading and receding travelling waves, as well as for rapidly receding travelling waves.  These approximations allow us to determine the relationship between $c$ and $\kappa$ so that commonly--reported experimental estimates of $c$ can be used to provide estimates of the unknown parameter $\kappa$.  Interestingly, when we reinterpret the Fisher--KPP model as a moving boundary problem, many disregarded features of the classical Fisher--KPP phase plane take on a new interpretation since travelling waves solutions with $c < 2$ are normally disregarded.  This 
\end{abstract}


\newpage
\section{Introduction} \label{intro}
Biological invasion is normally associated with situations where individuals within a population undergo both movement and proliferation events~\cite{Edelstein2005,Kot2003,Murray2002}. Such proliferation and movement, combined, can give rise to an \textit{invading front}. An invading front involves a population moving into a previously unoccupied space.  Ecologists are particularly interested in biological invasion.  For example, Skellam's~\cite{Skellam1951} work studies the invasion of muskrats in Europe; similarly, Otto and coworkers~\cite{Otto2018} study the spatial spreading of insects, whereas Bate and Hilker~\cite{Bate2019} study the invasion of predators in a predator--prey system.  As with many other similar examples, these three studies all make use of partial differential equation (PDE) models of invasion.

\begin{figure}[h]
	\centering
	\includegraphics[width=1\textwidth]{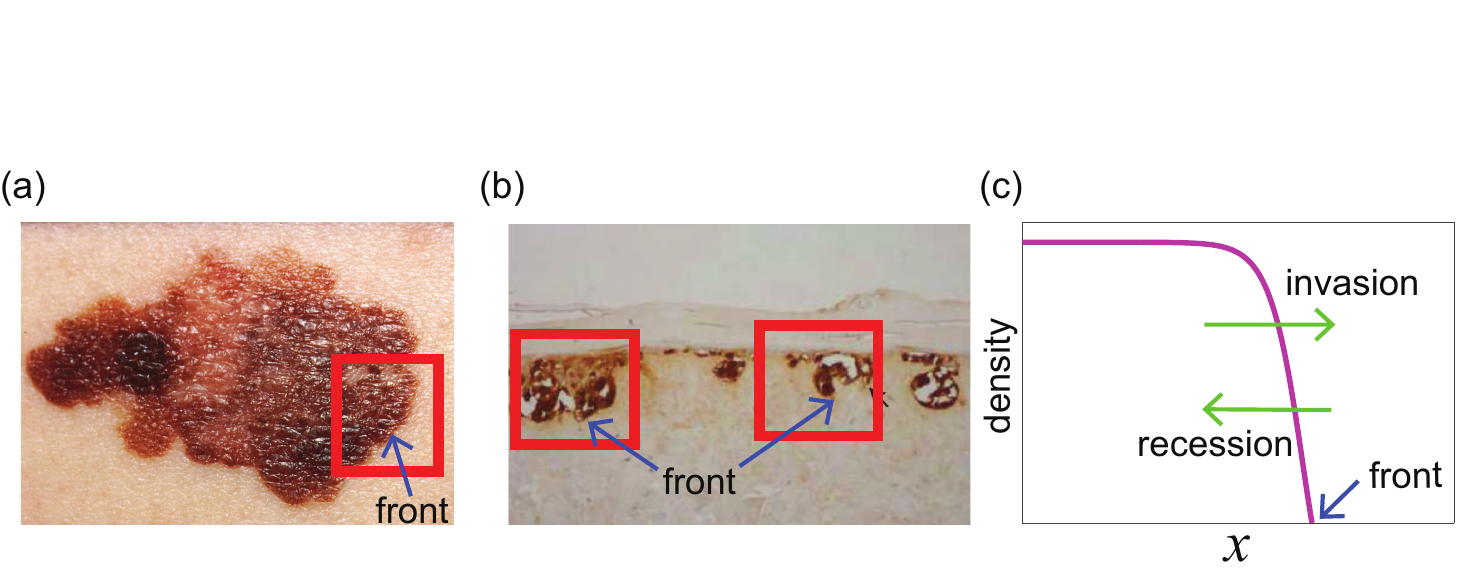}
	\caption{\textbf{Biological motivation}.  (a) Malignant melanoma (dark) spreading superficially across the skin surface~\cite{NCI1985} (reproduced with permission).  (b) Vertical cross section through a human skin equivalent experiment showing the inward invasion of a population of melanoma cells (dark)~\cite{Haridas2017,Haridas2018} (reproduced with permission). In (a)--(b) the region containing the leading edge of the invading population is highlighted in a red rectangle and the location of the sharp front is highlighted with blue arrows.  (c) Schematic solution of a mathematical model showing a sharp--fronted density profile that could either invade or recede, by moving in the positive or negative $x$--direction, respectively. In the schematic the location of the sharp front is also highlighted with a blue arrow.}
	\label{fig:1}
\end{figure}

Another common application of biological invasion is the study of cell invasion, including wound healing and malignant spreading.  Mathematical models of wound healing often consider the closure of a wound space by populations of cells that are both migratory and proliferative~\cite{Flegg2020,Jin2016,Jin2017,Maini2004a,Sherratt1990}. Malignant invasion involves combined migration and proliferation of tumour cells, which leads to tumour invasion into surrounding tissues~\cite{Byrne2010,Curtin2020,Strobl2020,Swanson2003}, as illustrated in Figure \ref{fig:1}(a)--(b), which shows the invasion of malignant melanoma cells.  Regardless of the application, many mathematical models of biological invasion involve the study of moving fronts, shown schematically in Figure \ref{fig:1}(c), using PDE models~\cite{Browning2019,Sengers2007,Warne2019}.  We interpret the schematic in Figure \ref{fig:1}(c) by thinking of the population as being composed of individuals that undergo diffusive migration with diffusivity $D > 0$, and logistic proliferation, with proliferation rate $\lambda > 0$.  As indicated, these two processes can lead to the spatial expansion as the population density profile moves in the positive $x$--direction.

The Fisher--KPP model~\cite{Canosa1973,Fisher1937,Kolmogorov1937,Edelstein2005,Murray2002} is probably the most commonly used reaction--diffusion equation to describe biological invasion in a single homogeneous population.  The Fisher--KPP model assumes that individuals in the population proliferate logistically and move according to a linear diffusion mechanism~\cite{Fisher1937,Kolmogorov1937}.  Travelling wave solutions of the Fisher--KPP model are often used to mimic biological invasion~\cite{Maini2004a,Maini2004b,Simpson2013}.  Long time solutions of the Fisher--KPP model that evolve from initial conditions with compact support eventually form smooth travelling waves without compact support such that $u(x,t) \to 0$ as $x \to \infty$.  These travelling wave solutions of the Fisher-KPP model move with speed $c = 2\sqrt{\lambda D}$~\cite{Edelstein2005,Murray2002}.  There are many other popular choices of single--species mathematical models of biological invasion;  for example, the \textit{Porous--Fisher} model~\cite{McCue2019,Sanchez1995,Sherratt1996,Simpson2011,Witelski1995} is a generalisation of the Fisher--KPP model with a degenerate nonlinear diffusion term which results in sharp--fronted travelling wave solutions.  Long time solutions of the Porous--Fisher model that evolve from initial conditions with compact support lead to invasion waves that move with speed $c = \sqrt{(\lambda D) /2}$~\cite{Murray2002}. Another generalisation of the Fisher--KPP model is the \textit{Fisher--Stefan} model~\cite{Du2010,Du2014a,Du2014b,Du2015}.  This approach involves reformulating the Fisher--KPP model as a moving boundary problem on $0 < x < L(t)$.  Setting the density to zero at the moving front, $x = L(t)$, means that the Fisher--Stefan model also gives rise to sharp--fronted solutions like the Porous--Fisher model~\cite{Elhachem2019}.  The motion of $L(t)$ in the Fisher--Stefan model is controlled by a one--phase Stefan condition~\cite{Crank1987,Dalwadi2020,Hill1987,Mitchell2014} with parameter $\kappa$.

Populations of motile and proliferative individuals do not always invade new territory; in fact, sometimes motile and proliferative populations recede or retreat.  The spatial recession of biological populations are often described in ecology.  For example, populations of desert locusts~\cite{Ibrahim2000}, plants in grazed prairies~\cite{Sinkins2012}, Arctic foxes~\cite{Killengreen2007} and dung beetles~\cite{Horgan2009} have all been observed to undergo both invasion and recession in different circumstances. While some previous mathematical models of biological invasion and recession have been described~\cite{Chaplain2020,Elhachem2020,Painter2003}, these previous models often focus on describing interactions between multiple subpopulations in a heterogeneous community rather than classical single species models, such as the Fisher--KPP model.  In fact, none of the three commonly--used single species models described here, the Fisher--KPP, Porous--Fisher or Fisher--Stefan models, have been used to study biological recession.  This is probably because neither the classical Fisher--KPP or Porous--Fisher models ever give rise to receding populations.  Given that the recession of population fronts is often observed, this limitation of the commonly--used Fisher--KPP and Porous--Fisher models is important and often overlooked.

\begin{figure}[h]
	\centering
	\includegraphics[width=1\textwidth]{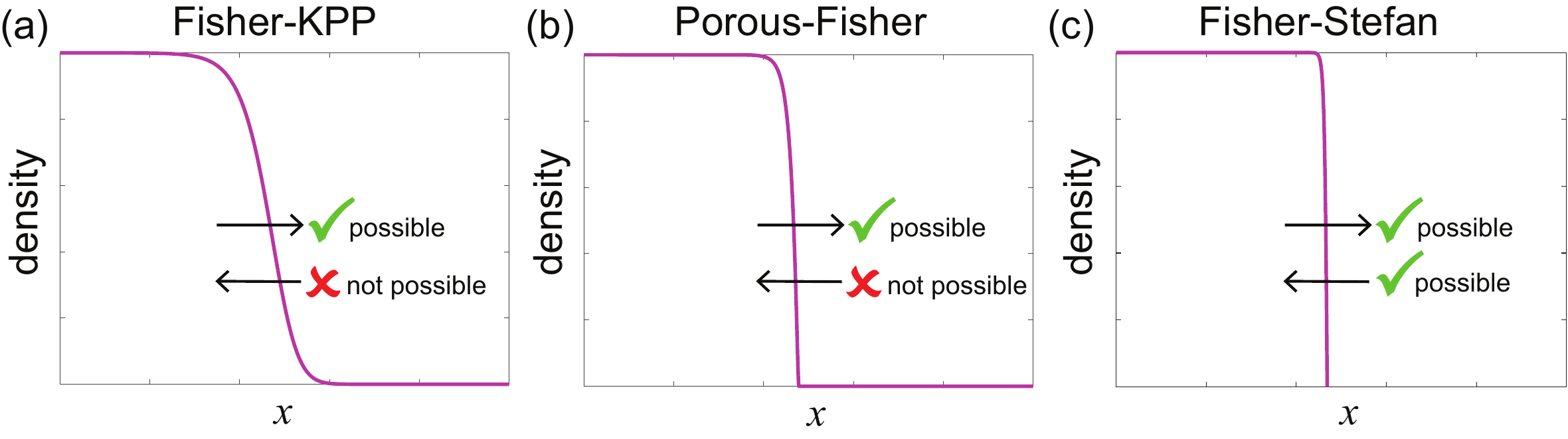}
	\caption{\cbl \textbf{Travelling wave schematic}.  (a) Travelling wave solution of the Fisher-KPP model supports invasion but not recession.  (b) Travelling wave solution of the Porous--Fisher model supports invasion but not recession.  (c) Travelling wave solution of the Fisher--Stefan model supports invasion and recession.\cb}
	\label{fig:2}
\end{figure}

\cbl  The ability of these three single--species models to support invading or receding travelling wave solutions is illustrated schematically in Figure \ref{fig:2}.  At this point it useful to provide a physical interpretation of what we mean by the \textit{invading} travelling wave.  If we consider a fixed position, $x=X$, a monotone invading travelling wave means that the density at that point, $u(X,t)$, increases with time, $\partial u(X,t) / \partial t > 0$.  In contrast, a monotone \textit{receding} travelling wave leads to the opposite behaviour where $\partial u(X,t) / \partial t < 0$ at a fixed position $x=X$.  This simple interpretation is useful because it holds regardless of the spatial orientation of the travelling wave.  For example, in this work we always consider moving fronts with the spatial orientation shown in Figure \ref{fig:1}(c).  Here, invasion is associated with movement in the positive $x$--direction and recession is associated with movement in the negative $x$--direction.  All results and definitions in this work hold when we consider fronts with the opposite spatial orientation, where invasion is associated with movement in the negative $x$--direction, and recession is associated with movement in the positive $x$--direction.  For convenience we adopt the usual convention shown in Figure \ref{fig:1}(c), but it is useful to remember that all results hold for the travelling waves with the opposite spatial orientation. \cb

In this work, we focus on the Fisher--Stefan model to study biological invasion \textit{and} recession. As just mentioned, unlike the classical Fisher--KPP and Porous--Fisher models, the Fisher--Stefan model can be used to simulate both biological invasion and recession.  One way of interpreting this difference is that the Fisher--Stefan model could be thought of as being more versatile than the more commonly--used Fisher--KPP or Porous--Fisher models.  As we will show, travelling wave solutions of the Fisher--Stefan model can be used to represent biological invasion with a positive travelling wave speed, $c >0$, as well as being able to model biological recession with a negative travelling wave speed, $c < 0$.  We explore these travelling wave solutions using full time--dependent numerical solutions of the governing PDE, phase plane analysis, and perturbation approximations.  A regular perturbation approximation around $c=0$ provides insight into both slowly invading and receding travelling waves, whereas a matched asymptotic expansion in the limit as $c \to -\infty$ provides insight into rapidly receding waves.  These perturbation solutions provide simple relationships between $\kappa$ and $c$.  \cbl For example, we show that slowly invading or receding travelling wave solutions of the Fisher-Stefan model move with speed $c \sim \kappa / \sqrt{3}$ as $\kappa \to 0$, whereas rapidly receding travelling wave solutions of the Fisher-Stefan model move with speed $c \sim 2^{-1} (\kappa+1)^{-1/2}$ as $\kappa \to -1^+$. \cb Such relationships are useful because estimates of $\kappa$ are not available in the literature, whereas experimental measurements of $c$ are relatively straightforward to obtain~\cite{Maini2004a,Maini2004b,Simpson2007}.

\section{Mathematical model} \label{sec:ModelFormulation}
In this work all dimensional variables and parameters are denoted with a circumflex and nondimensional quantities are denoted using regular symbols.  The Fisher--Stefan model is a reformulation of the classical Fisher--KPP equation to include a moving boundary,
\begin{equation}\label{eq:StefanFisher}
\dfrac{\partial \hat{u}}{\partial \hat{t}} = \hat{D} \dfrac{\partial^2 \hat{u}}{\partial \hat{x}^2} + \hat{\lambda} \hat{u}\left(1-\dfrac{\hat{u}}{\hat{K}}\right), \quad 0 < \hat{x} < \hat{L}(\hat{t}), \\
\end{equation}
where $\hat{u}(\hat{x},\hat{t}) \ge 0$ is the population density that depends upon position, $\hat{x}$, and time, $\hat{t} > 0$.  Individuals in the population move according to a linear diffusion mechanism with diffusivity $\hat{D} > 0$, the proliferation rate is $\hat{\lambda}>0$ and the carrying capacity density is $\hat{K}>0$.

We consider the Fisher--Stefan model on $0 < \hat{x} < \hat{L}(\hat{t})$, with a zero flux condition at the origin. The sharp front is modelled by setting the density to be zero at the leading edge, giving
\begin{equation}
\label{eq:BCNeumannDirichlet}
\dfrac{\partial \hat{u}(0,\hat{t})}{\partial \hat{x}}  = 0, \qquad\qquad \hat{u}(\hat{L}(\hat{t}),\hat{t}) = 0.
\end{equation}
The evolution of the domain is controlled by a classical one--phase Stefan condition that relates the speed of the moving front to the spatial gradient of the density profile at the moving boundary,
\begin{equation}
\label{eq:BCStefan}
\dfrac{\text{d}\hat{L}(\hat{t})}{\text{d} \hat{t}} = -\left.{\hat{\kappa} \frac{\partial \hat{u}}{\partial \hat{x}}}\right|_{\hat{x}=\hat{L}(\hat{t})},
\end{equation}
where $\hat{\kappa}$ is a constant to be specified~\cite{Crank1987,Dalwadi2020,Hill1987,Mitchell2014}.  While it is possible to consider different, potentially more complicated conditions at the moving boundary~\cite{Crank1987,Elhachem2020,Gaffney1999,Hill1987}, here we restrict our attention to the classical one--phase Stefan condition.

In the context of cell invasion, typical values of $\hat{D}$ are approximately 100--3000 $\mu$m$^2$/h~\cite{Johnston2015,Johnston2016,Jin2016}; typical values  $\hat{\lambda}$ are approximately 0.04--0.06 /h~\cite{Johnston2015,Jin2016}; and typical values of the carrying capacity density are 0.001--0.003 cells/$\mu$m$^2$~\cite{Johnston2015,Jin2016}.   To simplify our analysis we will now nondimensionalise the Fisher--Stefan model.

\subsection{Nondimensional model} \label{sec:NomDimModel}
Introducing dimensionless variables, $x = \hat{x}\sqrt{\hat{\lambda}/\hat{D}}$, $t = \hat{\lambda} \hat{t}$, $u = \hat{u}/\hat{K}$, $L(t) = \hat{L}(\hat{t})\sqrt{\hat{\lambda}/\hat{D}}$ and $\kappa = \hat{\kappa}/\hat{D}$, the Fisher--Stefan model can be simplified to give
\begin{align}
&\dfrac{\partial u}{\partial t} =  \dfrac{\partial^2 u}{\partial x^2} +  u\left(1-u \right), \quad 0<x<L(t), \label{eq:NondimPDE} \\
&\dfrac{\partial u(0,t)}{\partial x} = 0, \qquad\qquad u(L(t),t) = 0, \label{eq:NondimBCDirichletNeumann}\\
&\dfrac{\text{d}L(t)}{\text{d} t} = -\kappa \frac{\partial u(L(t),t)}{\partial x}, \label{eq:NondimBCStefan}
\end{align}
so that we only need to specify one parameter, $\kappa$ together with initial conditions for $u$ and $L$.  As mentioned previously, estimates of diffusivity, proliferation rate and carrying capacity in the context of cell invasion are available in the literature~\cite{Jin2016,Maini2004a}.  In contrast, estimates of $\kappa$ are not.  Therefore, one of the aims of this work is to provide mathematical insight into how estimates of $\kappa$ can be obtained, and we will provide more discussion on this point later.

In all cases where we consider time--dependent solutions of Equations (\ref{eq:NondimPDE})--(\ref{eq:NondimBCStefan}) we always choose the initial condition to be
\begin{equation}\label{eq:IC}
u(x,0) = \alpha \left(1 - \mathrm{H}[L(0)]\right),
\end{equation}
where $\alpha>0$ is a positive constant and $\mathrm{H}[\cdot]$ is the Heaviside function, so that $u(x,0)=\alpha$ for $x < L(0)$ and $u(x,L(0))=0$.

To solve Equations (\ref{eq:NondimPDE})--(\ref{eq:IC}) numerically, we transform the governing equations from an evolving domain, $0 < x < L(t)$ to a fixed domain, $0 < \xi < 1$ by setting $\xi = x/L(t)$. The transformed equations on the fixed domain are spatially discretised using a uniform finite difference mesh and standard central finite difference approximations.  The resulting system of nonlinear ordinary differential equations (ODE) is integrated through time using an implicit Euler approximation. Newton--Raphson iteration and the Thomas algorithm are used to solve the resulting system of nonlinear algebraic equations~\cite{Simpson2005}.  Full details of the numerical method are given in the Supplementary Material; MATLAB implementation of the algorithm is available on \href{https://github.com/maudelhachem/El-Hachem2020b}{GitHub}.

\section{Results and Discussion} \label{sec:Results}
We begin our analysis of the Fisher--Stefan model by presenting some time--dependent solutions of Equations (\ref{eq:NondimPDE})--(\ref{eq:IC}) before analysing these solutions using the phase plane and perturbation techniques.

\subsection{Time--dependent partial differential equation solutions} \label{sec:NumSims}
\begin{figure}[H]
	\centering
	\includegraphics[width=0.7\linewidth]{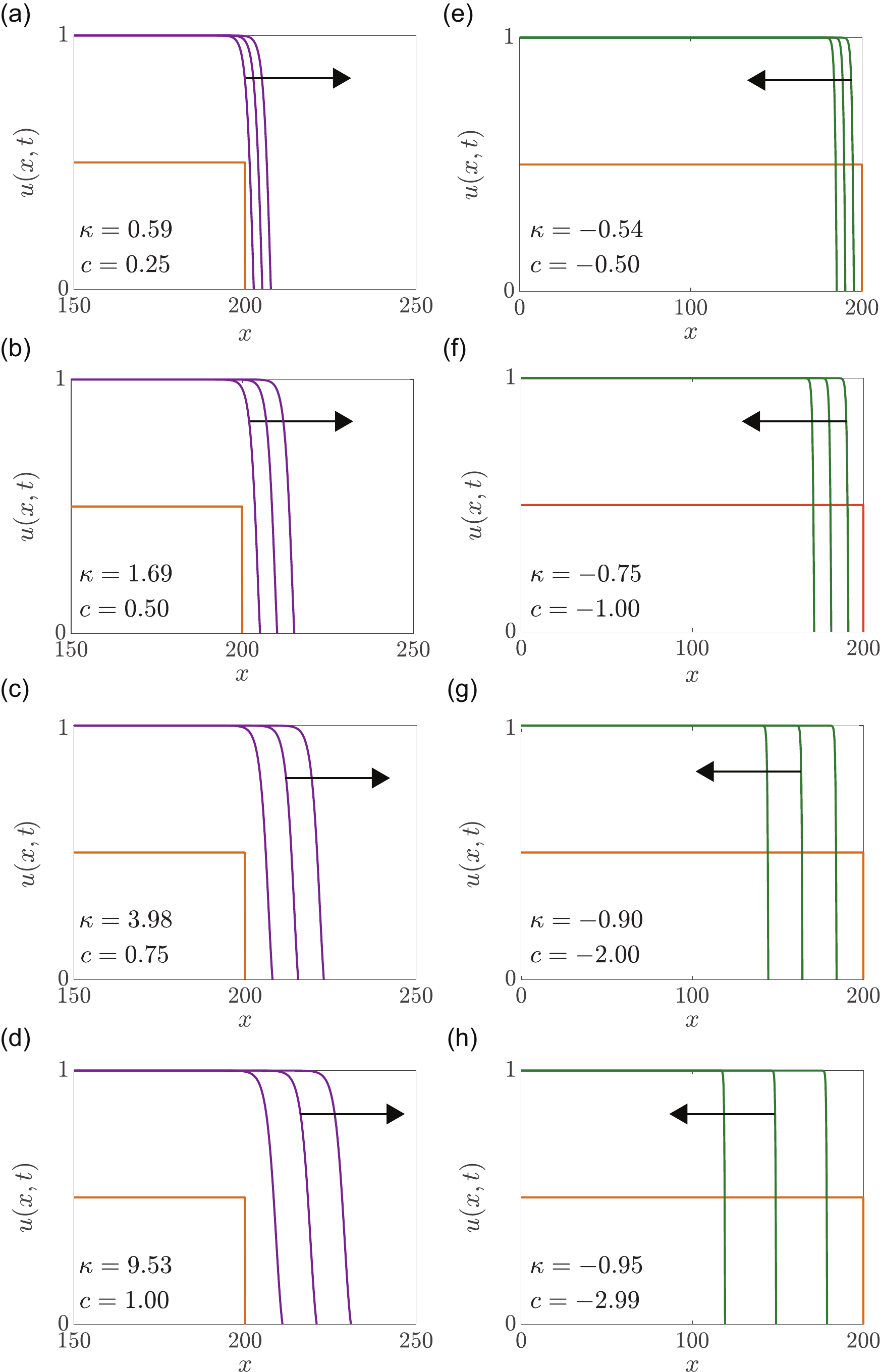}
	\caption{\textbf{Invading and receding travelling wave solutions of the Fisher--Stefan model.}  Numerical solutions of Equations (\ref{eq:NondimPDE})--(\ref{eq:IC}) are given at $t = 0,10,20$ and $30$. The initial condition is given by Equation (\ref{eq:IC}) with $\alpha=0.5$ and $L(0)=200$.  Results in (a)--(d) lead to invading travelling waves with $c=0.25,0.50,0.75$ and $1.00$, respectively.  These travelling waves are obtained by choosing $\kappa = 0.5859, 1.6879, 3.9823$ and $9.5315$, respectively.  Results in (e)--(h) lead to receding travelling waves with $c=-0.50,-1.00,-2.00$ and $-2.99$, respectively. These receding travelling waves are obtained by choosing $\kappa = -0.5387, -0.7529, -0.9036$ and $-0.9510$, respectively.  Our estimates of $c$ correspond are obtained at late time, here $t=30$.  Note that estimates of $\kappa$ are reported in the caption to four decimal places, whereas the estimates given in the subfigures are reported to two decimal places to keep the figure neat.}
	\label{fig:3}
\end{figure}

Results in Figure \ref{fig:3} show a suite of numerical solutions of Equations (\ref{eq:NondimPDE})--(\ref{eq:IC}) plotted at regular time intervals.  Similar to our previous work~\cite{Elhachem2019}, the results in Figure \ref{fig:2}(a)--(d) suggest that the initial condition evolves into invading travelling waves for $\kappa > 0$.  However, unlike our previous work, the results in Figure \ref{fig:2}(e)--(h) show that we obtain receding travelling waves for $\kappa < 0$.  To obtain these solutions we specify a value of $\kappa$, as indicated in each subfigure, and then measure the eventual speed of the travelling wave, $c$, by estimating $\textrm{d} L(t) / \textrm{d} t$ using the numerical solution of the PDE as described in the Supplementary Material.  Therefore, in this approach to studying the travelling wave solutions, we treat $\kappa$ as an input to the numerical algorithm, and $c$ is an output.  In fact, in generating results in Figure \ref{fig:2} we took great care to choose $\kappa$ so that our resulting estimates of $c$ are clean values, such as $c=0.25, 0.50, 0.75$ and $1.00$.  We will explain how to make this choice later, in Section \ref{sec:PhasePlane}.

\cbl All results in Figure \ref{fig:3} correspond to the initial condition (\ref{eq:IC}) with $\alpha = 0.5$.  Additional results in the Supplementary Material show similar results for different initial conditions by varying the choice of $\alpha = 0.25, 0.75$ and 1.00.  These additional results strongly suggest that the time-dependent solutions of  Equations (\ref{eq:NondimPDE})--(\ref{eq:IC}) always approaches the same travelling wave solution with the same speed, $c$, regardless of the choice of $\alpha$.  \cb

Results in Figure \ref{fig:3}  show that $c$ is an increasing function of $\kappa$. The density profile at the leading edge is sharp in all cases and indeed the slope of $u$ at $x=L(t)$ decreases as $\kappa$ decreases.  The shape of the density profile differs depending on whether we consider an invading or receding travelling wave, since the receding travelling waves are much steeper than the invading travelling waves.  These numerical results in Figure \ref{fig:2}  are interesting since neither the Fisher--KPP nor the Porous--Fisher can be used to simulate this range of behaviours.  The feature of the Fisher--Stefan model which enables us to simulate both invasion and retreat is the choice of $\kappa$.  We will now explore the relationship between $c$ and $\kappa$ by studying the travelling wave solutions in the phase plane.

\cbl Interpreting the Stefan condition, Equation (\ref{eq:NondimBCStefan}), in terms of the underlying biology is an open question that is very interesting.   In essence, the Stefan condition states that the time rate of change of the right-most position of the boundary is proportional to the spatial gradient of the density at that point, $\textrm{d} L(t) / \textrm{d}t \propto \partial u(L(t),t) / \partial x$.  There are many ways to interpret this widely--used boundary condition.  In the usual geometry, shown in Figure \ref{fig:1}(a), we have $\partial u(L(t),t) / \partial x < 0$, and setting the coefficient of proportionality to be negative leads to the standard case where $L(t)$ increases.  One way of interpreting this is that the position of the boundary evolves so that $L(t)$ moves down the spatial gradient of $u(x,t)$ at $x=L(t)$.  In the same situation as in Figure \ref{fig:1}(a), where $\partial u(L(t),t) / \partial x < 0$, setting the coefficient of proportionality to be positive leads to $L(t)$ decreasing.  One way of interpreting this is that the position of the boundary evolves so that $L(t)$ moves up the spatial gradient of $u(x,t)$ at $x=L(t)$. Of course, this theoretical interpretation is not tested or confirmed biologically, but this distinction between invasion and recession, dictated by the sign of the proportionality coefficient in the Stefan condition, is analogous to the distinction between \textit{chemoattraction} and \textit{chemorepulsion} in bacterial and cellular chemotaxis~\cite{Edelstein2005,Keller1971,Murray2002}.  In practical terms we provide a description of how $\kappa$ could be estimated using simple experiments in the Discussion section. \cb

\subsection{Phase plane analysis} \label{sec:PhasePlane}
To analyse travelling wave solutions of the Fisher--Stefan model in the phase plane we consider Equation (\ref{eq:NondimPDE}) in terms of the travelling wave coordinate, $z = x-ct$ and we seek solutions of the form $u(x,t)=U(z)$ which leads to the following ODE,
\begin{equation} \label{eq:ODEUz}
\frac{\mathrm{d}^2 U}{\mathrm{d} z^2} + c\frac{\mathrm{d} U}{\mathrm{d} z} + U(1-U) = 0,  \quad  -\infty < z < 0,
\end{equation}
with boundary conditions
\begin{align}
U(-\infty) &= 1, \quad   U(0)= 0, \label{eq:ODEU_BC} \\
c &= -\kappa \frac{\mathrm{d} U(0)}{\mathrm{d} z}, \label{eq:ODEStefan}
\end{align}
where we choose $z=0$ to correspond to the moving boundary.

To study Equation (\ref{eq:ODEUz}) in the phase plane we rewrite this second order ODE as a first order dynamical system
\begin{align}
\frac{\text{d}U}{\text{d} z} & = V, \label{eq:ODEdU}\\
\frac{\text{d}V}{\text{d} z} &= -cV - U(1-U), \label{eq:ODEdV}
\end{align}
with the equilibrium points $(0,0)$ and $(1,0)$.  Equations (\ref{eq:ODEdU})--(\ref{eq:ODEdV}) are the well--known dynamical system  associated with travelling wave solutions of the classical Fisher--KPP model~\cite{Canosa1973,Edelstein2005,Murray2002}.  Therefore, many previous results for this system also apply here to the Fisher--Stefan model.  For example, linear stability analysis shows that $(1,0)$ is a saddle point for all values of $c$, whereas $(0,0)$ is a stable node if $c \ge 2$; a stable spiral if $0 < c < 2$; a centre if $c=0$; an unstable spiral if $-2 < c < 0$; and, an unstable node if $c \le -2$.  Typically, in the regular analysis of the Fisher--KPP model the possibility of travelling wave solutions with $c < 0$ (and $\partial u / \partial x < 0$) is never considered because time--dependent numerical solutions of the Fisher--KPP model only ever evolve into invading travelling waves with positive wave speed.  Further, in the regular analysis of the Fisher--KPP model, the possibility of travelling waves with $c < 2$ is disregarded because linear stability analysis shows that $(0,0)$ is a stable spiral, implying that $U(z) < 0$ for various intervals in $z$~\cite{Murray2002}.  Our previous work has shown that this caution is not required for the Fisher--Stefan model as these often--neglected trajectories in the phase plane are, in fact, associated with physically--relevant travelling wave solutions~\cite{Elhachem2019}.

\begin{figure}[H]
	\centering
	\includegraphics[width=0.7\linewidth]{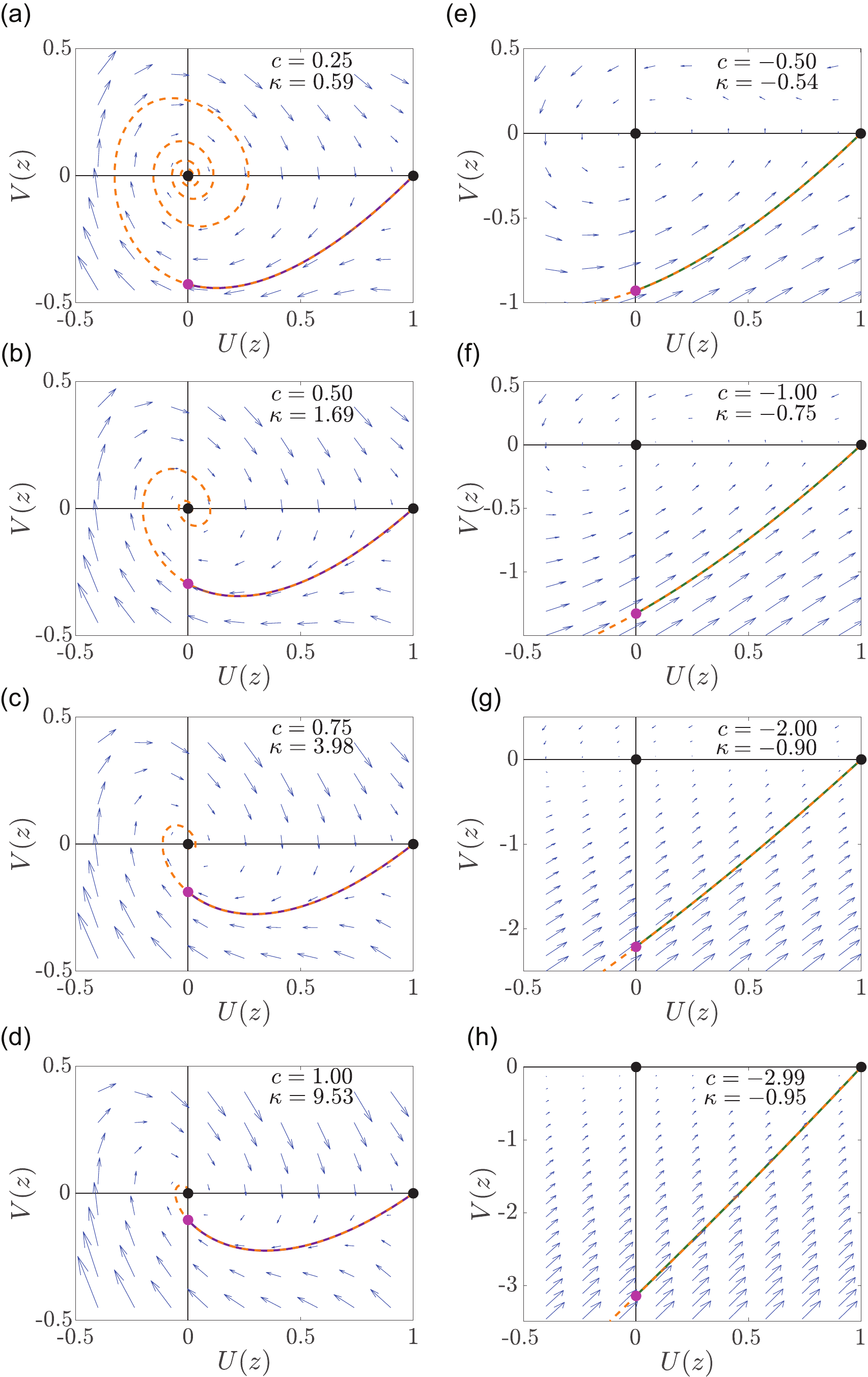}
	\caption{\textbf{Phase plane for travelling wave solutions of the Fisher--Stefan model.} Equilibrium points are shown as black discs, and the point at which the trajectory intersects the $V(z)$ axis are shown as pink discs. The numerical solution of the dynamical system, Equations (\ref{eq:ODEdU})--(\ref{eq:ODEdV}) is shown in dashed orange and the travelling wave solution obtained from the numerical time--dependent PDE solutions, Equations (\ref{eq:NondimPDE})--(\ref{eq:IC}) is superimposed in solid purple for the invading travelling waves in (a)--(d) and in solid green for the receding travelling waves in (e)--(h).  The flow associated with the dynamical system is shown with blue vectors obtained using Matlab's quiver function.}
	\label{fig:4}
\end{figure}

To explore these ideas will now visualise the phase plane for each travelling wave shown previously in Figure \ref{fig:3}.  To show trajectories in the phase plane we solve Equations (\ref{eq:ODEdU})--(\ref{eq:ODEdV}) numerically using Heun's method.  A Matlab implementation of our algorithm to visualise these phase planes is available on \href{https://github.com/maudelhachem/El-Hachem2020b}{GitHub}.  Unlike the full time--dependent solution of the PDE model where we treat $\kappa$ as the input and $c$ as the output of the numerical algorithm, here in the phase plane we treat $c$ as the input into the numerical algorithm to generate the phase plane trajectory and we use this trajectory to estimate  $\kappa$, as we will now explain.  Phase planes for $c = 0.25,0.50,0.75$ and $1.00$ are given in Figure \ref{fig:4}(a)--(d), respectively.  Similarly, phase planes for $c = -0.50,-1.00,-2.00$ and $-2.99$ are given in Figure \ref{fig:4}(e)--(f), respectively.    Each phase plane in Figure \ref{fig:4} corresponds to the particular PDE solution in Figure \ref{fig:3}.

The phase planes in Figure \ref{fig:4}(a)--(d) correspond to invading fronts with various values of $0<c<2$. As we previously describe~\cite{Elhachem2019}, these phase plane trajectories are usually neglected in the usual analysis of the Fisher--KPP model since they leave near $(1,0)$ and eventually spiral into $(0,0)$ as $z \to \infty$, implying that $U(z)<0$ for certain intervals along the trajectory.  In contrast, the travelling wave solution of the Fisher--Stefan model must also satisfy the Stefan condition at $U(z)=0$, which means that we truncate the trajectory at $z=0$ and only focus on that part of the trajectory in the fourth quadrant of the phase plane where $U(z) > 0$.  Each trajectory in Figure \ref{fig:4}(a)--(d) intersects the $V(z)$ axis at a special point, $(0,V^*)$, which corresponds to the Stefan condition where $U=0$ and $c = -\kappa V^*$.  Estimating $V^*$ from the numerically--generated phase plane trajectory allows us to estimate $\kappa$.  Following this approach we obtain estimates of $\kappa$ for each value of $c$, and these estimates compare very well with the estimates used to generate the time--dependent PDE solutions in Figure \ref{fig:3}.  These phase planes explain why invading travelling waves for the Fisher--Stefan model are restricted to $0 < c < 2$ since setting $c > 2$ means that the origin is a stable node and the heteroclinic orbit between $(1,0)$ and $(0,0)$ never intersects the $V(z)$ axis, giving  $c \to 2^-$ as $\kappa \to \infty$~\cite{Du2010,Elhachem2019}.

For completeness we also show the remaining portion of the phase plane trajectory in Figure \ref{fig:4}(a)--(d) that eventually spirals into $(0,0)$ as $z \to \infty$. Further, for each phase plane in Figure \ref{fig:3}(a)--(d) we take the late time PDE solution from Figure \ref{fig:3}(a)--(d) and transform these PDE solutions into a $(U(z),V(z))$ phase plane trajectory, and superimpose these curves in the phase planes in Figure \ref{fig:4}(a)--(d).  In each case the trajectory obtained by solving the dynamical system numerically is visually indistinguishable, at this scale, from the trajectory obtained by plotting the PDE solutions in the phase plane.

The phase planes in Figure \ref{fig:4}(e)--(h) correspond to receding travelling waves with various $c<0$.  As we previously describe, these phase planes for $c < 0$ are not normally considered for the Fisher--KPP model since receding travelling wave solutions of the Fisher-KPP model are not possible.  Here we see that we are interested in that part of the trajectory in the fourth quadrant that leaves $(0,V^*)$ and joins $(1,0)$ as $z \to \infty$.  Again, we can use this trajectory to estimate $\kappa$ and the estimates from the phase plane compare well with the values used in the full time--dependent PDE solutions in Figure \ref{fig:3}(e)--(h).  For completeness we take the late--time PDE solutions in Figure \ref{fig:3}(e)--(h) and superimpose these trajectories in Figure \ref{fig:4}(e)--(h) where we see that the numerical solution of the trajectory obtained from the dynamical system is again visually indistinguishable from the trajectory obtained from the PDE solutions.  Unlike the invading travelling wave solutions where linear stability analysis in the phase plane gives us the condition that $0 < c < 2$, there is no restriction on $c$ in the phase plane so that the Fisher--Stefan model gives rise to receding travelling waves with $-\infty < c < 0$.

Now we have shown that both invading and receding travelling wave solutions of the Fisher--Stefan model can be studied in the phase plane, we will analyse the governing equations in the phase plane to provide more detailed insight into the relationship between $\kappa$ and $c$.  This will be important because estimates of $\kappa$ are not available in the literature, whereas estimates of $c$ are easier to obtain experimentally~\cite{Maini2004a,Maini2004b,Simpson2007}.

\subsection{Analysis} \label{sec:Analysis}
\subsubsection{Exact solution for stationary waves} \label{sec:Exactsolution}
Here we solve for the shape of the stationary travelling wave when $c=0$ by re--writing Equations (\ref{eq:ODEdU})--(\ref{eq:ODEdV}) as
\begin{equation}\label{eq:odeVU}
\frac{\text{d}V}{\text{d}U} = \frac{-cV - U(1-U)}{V},
\end{equation}
where it is clear that an exact solution for $V(U)$ can be obtained when $c=0$. This solution can be written as
\begin{equation}
\label{eq:exactVU}
V(U) =  \pm \sqrt{-U^2+\dfrac{2U^3+1}{3}},
\end{equation}
where, we are primarily interested in the negative solution since $V < 0$ at the leading edge.  Equation (\ref{eq:exactVU}) with $U(0)=0$ can be integrated to  give the shape of the stationary wave,
\begin{equation}
\label{eq:exactUz}
U(z) =  \dfrac{3}{2} \left[\tanh \left(\dfrac{z}{2}-\arctanh\dfrac{\sqrt{3}}{3}\right)^2-1\right].
\end{equation}

\begin{figure}[H]
	\centering
	\includegraphics[width=\linewidth]{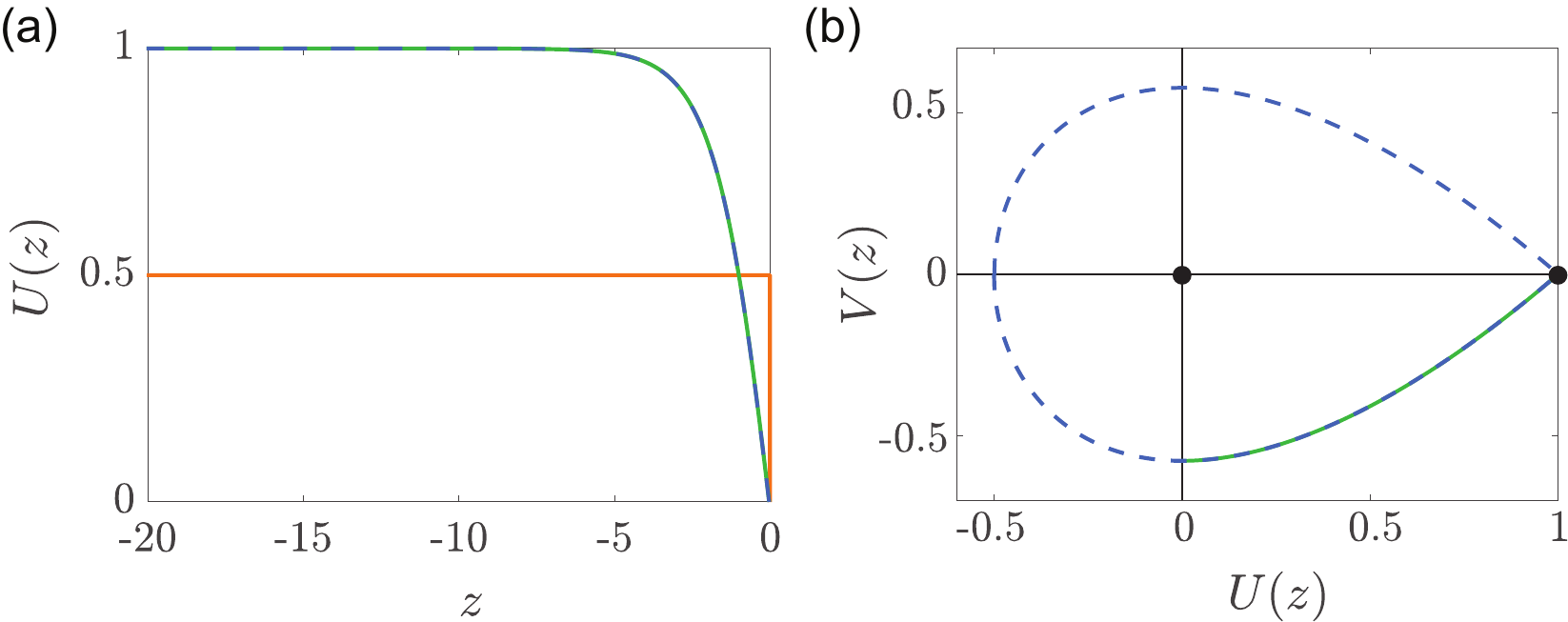}
	\caption{\textbf{Exact solution for the stationary travelling wave, $c=0$.} (a) Comparison of the exact solution, Equation (\ref{eq:exactUz}), in dashed blue with the numerical solution of Equations (\ref{eq:NondimPDE})--(\ref{eq:IC}) with $\kappa=0$ in solid green. The initial condition for the numerical solution of the PDE is in orange. (b) Comparison of the exact solution of the phase plane trajectory, Equation (\ref{eq:exactVU}), in dashed blue, with the trajectory obtained by plotting the PDE solution in the phase plane in solid green.  Equilibrium points in the phase plane are shown with black discs.}
	\label{fig:5}
\end{figure}

Results in Figure \ref{fig:5} compare these exact solutions for $c=0$ with various numerical solutions.  Firstly, in Figure \ref{fig:5}(a) we show a time--dependent solution of  Equations (\ref{eq:NondimPDE})--(\ref{eq:IC}) with $\kappa=0$ which evolves into a stationary wave that is visually indistinguishable from the exact solution, Equation (\ref{eq:exactUz}), at this scale.  The phase plane in Figure \ref{fig:5}(b) shows the late--time PDE solution from Figure \ref{fig:5}(a) plotted as a trajectory in the $(U(z),V(z))$ phase plane.  In this phase plane we superimpose the exact solution, Equation (\ref{eq:exactVU}), which forms a homoclinic orbit in the shape of a teardrop.  \cbl The part of the homoclinic orbit in the fourth quadrant of the phase plane corresponds to the stationary wave, and we see that the numerical trajectory and the exact solution are indistinguishable at this scale.  Just as we observed for the invading travelling waves in Figure \ref{fig:4}, the stationary wave here corresponds to just one part of a trajectory in the phase plane.  This is different to the usual phase plane analysis for either the Fisher-KPP or Porous--Fisher models where travelling wave solutions correspond to a complete trajectory, rather than just part of a trajectory. \cb

\subsubsection{Perturbation solution for slowly invading or receding travelling waves} \label{sec:RegularPerturbation}
Results in Section \ref{sec:Exactsolution} show that we have an exact solution when $c=0$.  We now seek a perturbation solution for $|c| \ll 1$ by writing~\cite{Murray1984},
\begin{align} \label{eq:Vregexpansion}
V(U) &= V_0(U) + c V_1(U)+ c^2 V_2(U) +\mathcal{O}(c^3).
\end{align}
Substituting Equation (\ref{eq:Vregexpansion}) into Equation (\ref{eq:odeVU}) gives,
\begin{align}
\label{eq:ODEV0}
&\frac{\text{d}V_0}{\text{d}U}V_0 + U(1-U) = 0,&V_0(1) = 0,\\
&\frac{\text{d}V_1}{\text{d}U}V_0 + \frac{\text{d}V_0}{\text{d}U}V_1 +V_0 = 0,&V_1(1) = 0, \label{eq:ODEV1}\\
&\frac{\text{d}V_2}{\text{d}U}V_0 + \frac{\text{d}V_0}{\text{d}U}V_2 + V_1\left(\frac{\text{d}V_1}{\text{d}U} +1\right) = 0,&V_2(1) = 0. \label{eq:ODEV2}
\end{align}
The solutions of these differential equations are
\begin{align}
\label{eq:V0csmall}
V_0(U) &= \dfrac{\sqrt{3(2U+1)}}{3}(U-1),\\
V_1(U) &= \dfrac{-(U-2)(1+2U)^{3/2}-3\sqrt{3}}{5(U-1)\sqrt{1+2U}}, \label{eq:V1csmall}\\
\begin{split}
	V_2(U) &=  \dfrac{-18\sqrt{3}}{25(2U+1)^{3/2}(U-1)(\sqrt{6U+3}-3)^2(\sqrt{6U+3}+3)^2} \\
	&\times\Bigg(-2U^3(6U^2-15U+20)+15U(U+2)+31\\
	&+\sqrt{6U+3}\left[(2U+1)(6U+3)-30U-15\right]\\
	&\left.+(60U^3-90U^2+30)\ln\left[\dfrac{(\sqrt{6U+3}+3)(U-1)}{6(\sqrt{6U+3}-3)}\right]\right). \label{eq:V2csmall}
\end{split}
\end{align}
Maple code to generate these solutions is available on \href{https://github.com/maudelhachem/El-Hachem2020b}{GitHub}.
These three solutions can be used to truncate Equation (\ref{eq:Vregexpansion}) at different orders, and in doing so we will make use of the $\mathcal{O}(1)$, $\mathcal{O}(c)$ and $\mathcal{O}(c^2)$ perturbation solutions.  Given our various approximate perturbation solutions for $V(U)$, we can either directly plot these solution in the phase plane and compare them with numerically--generated phase plane trajectories, or we can integrate these perturbation solutions numerically to give an approximation for the shape of the travelling wave, $U(z)$.  To estimate the shape of the travelling wave we integrate the perturbation solution for $V(U)$ using Heun's method with $U(0) = 0$, and we integrate from $z=0$ to $z=-Z$, where $Z$ is taken to be sufficiently large.

\begin{landscape}
	\begin{figure}[h]
		\centering
		\includegraphics[height=0.8\textheight]{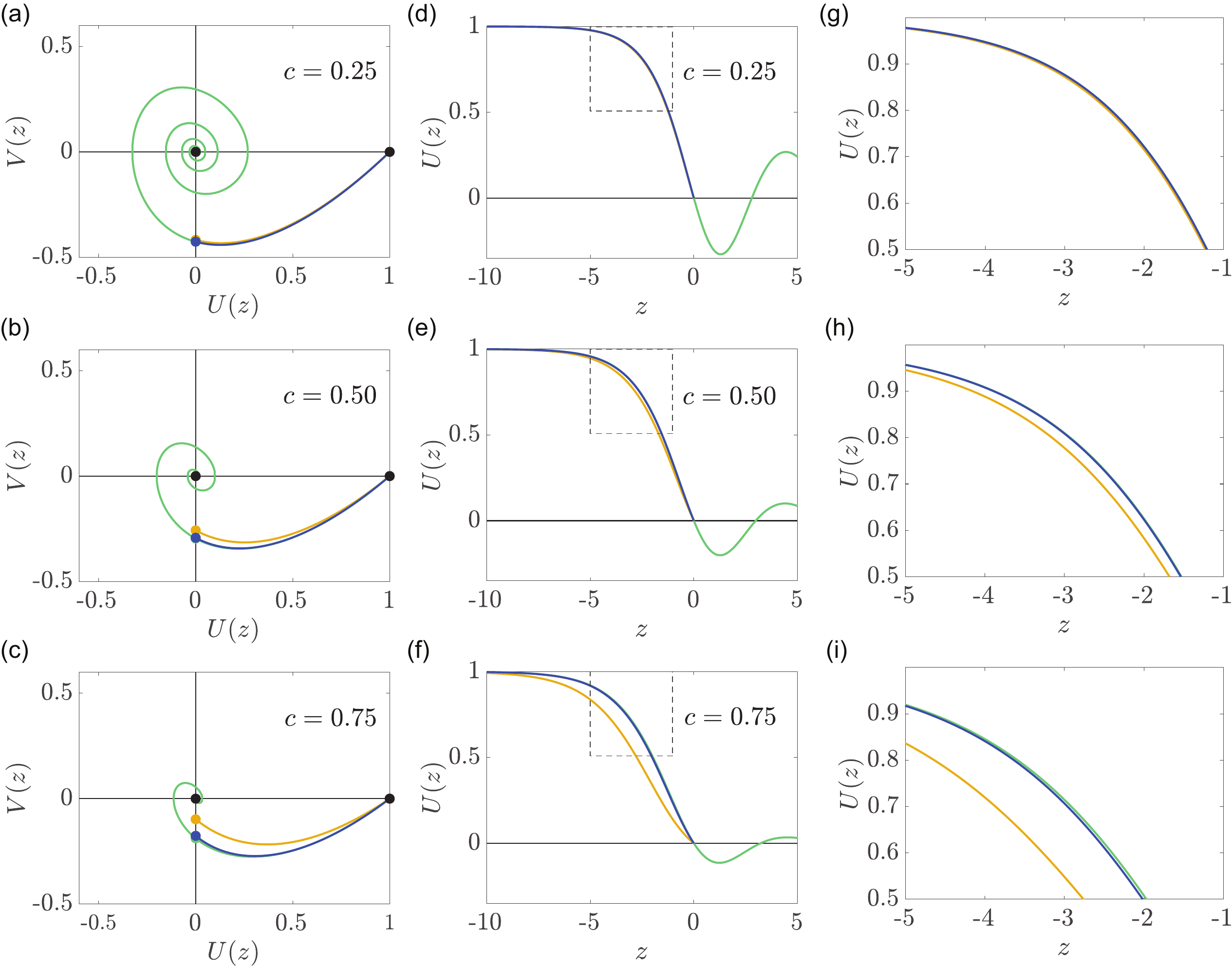}
		\caption{\textbf{Perturbation solutions for slow invading travelling waves.} (a)--(c) show the phase plane for $c=0.25, 0.50$ and 0.75, respectively.  Equilibrium points are shown with black discs.  The numerical solution of Equations (\ref{eq:ODEdU})--(\ref{eq:ODEdV}) are shown in green and the point at which these trajectories intersect the $V(z)$ axis are shown with a green disc.  The $\mathcal{O}(c)$ and $\mathcal{O}(c^2)$ perturbation solutions are shown in yellow and blue, respectively.  The intersection of the $V(z)$ for the  $\mathcal{O}(c)$ and $\mathcal{O}(c^2)$ perturbation solutions are shown in a yellow and blue disc, respectively. Results in (d)--(f) compare the shape of the travelling wave profile, $U(z)$, obtained using the numerical solution of the phase plane trajectory (green) with the  $\mathcal{O}(c)$ and $\mathcal{O}(c^2)$ perturbation solutions in yellow and blue, respectively.  Results in (g)--(i) show magnified comparison of the three solutions in the regions highlighted by the dashed boxes in (d)--(f).}
		\label{fig:6}
	\end{figure}
\end{landscape}

We now compare various perturbation solutions with phase plane trajectories and time--dependent PDE solutions for both invading and receding travelling waves.  Figure \ref{fig:6} focuses on invading travelling wave with $c > 0$.  Results in Figure \ref{fig:5}(a)--(c) show  the phase plane for $c=0.25, 0.50$ and $0.75$, respectively. The numerical solution of the dynamical system is shown in green, and is superimposed on the $\mathcal{O}(c)$ and $\mathcal{O}(c^2)$ perturbation solutions in yellow and blue, respectively.  In these results there is a visual difference between the numerically--generated phase plane trajectories and the $\mathcal{O}(c)$ perturbation solutions, however the $\mathcal{O}(c^2)$ perturbation solution compares very well with the numerically--generated phase plane trajectories.

Results in Figure \ref{fig:6}(d)--(f) compare the shape of the travelling wave, $U(z)$, using the numerical solution of the dynamical system in the phase plane with the results obtained from the $\mathcal{O}(c)$ and $\mathcal{O}(c^2)$ perturbation solutions.  For the numerical solution of the dynamical system we deliberately show the invasion profile using the trajectory from $z=-15$ to $z=5$, which includes the unphysical part of the trajectory, $z > 0$, where $U(z)$ is oscillatory.  To make a clear distinction between the physical and unphysical parts of the invading profile we include a horizontal line at $U(z) = 0$.  The horizontal line emphasise the fact that $U(z) > 0$ for $z < 0$, and $U(z)$ is oscillatory for $z > 0$. All three solutions are visually indistinguishable at the scale shown in Figure \ref{fig:6}(d) where $c=0.25$.  For $c=0.50$ and $c=0.75$ we see a  visually--distinct difference between the profiles from the phase plane trajectory and the $\mathcal{O}(c)$ perturbation solutions, whereas the $\mathcal{O}(c^2)$ perturbation solution gives an excellent approximation for these larger speeds.  Results in Figure \ref{fig:6}(g)--(i) show magnified comparisons of the shape of $U(z)$ corresponding to the dashed inset regions in Figure \ref{fig:6}(d)--(f) where it is easier to see the distinction between the three solutions.

\begin{landscape}
	\begin{figure}[h]
		\centering
		\includegraphics[height=0.8\textheight]{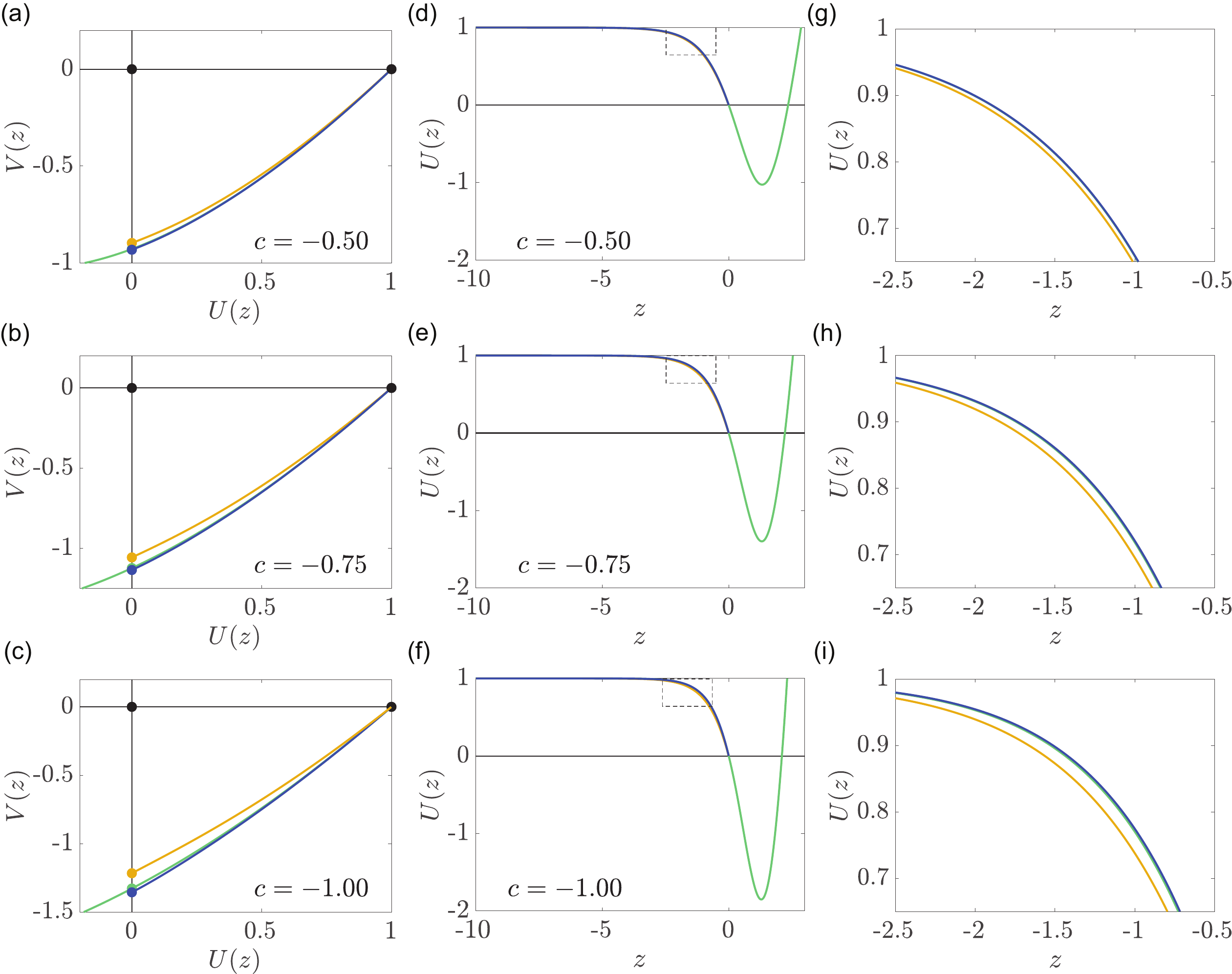}
		\caption{\textbf{Perturbation solutions for slow receding travelling waves.} (a)--(c) show the phase plane for $c=-0.50, -0.75$ and $-1.00$, respectively.  Equilibrium points are shown with black discs.  The numerical solution of Equations (\ref{eq:ODEdU})--(\ref{eq:ODEdV}) are shown in green and the point at which these trajectories intersect the $V(z)$ axis are shown with a green disc.  The $\mathcal{O}(c)$ and $\mathcal{O}(c^2)$ perturbation solutions are shown in yellow and blue, respectively.  The intersection of the $V(z)$ for the  $\mathcal{O}(c)$ and $\mathcal{O}(c^2)$ perturbation solutions are shown in a yellow and blue disc, respectively. Results in (d)--(f) compare the shape of the travelling wave profile, $U(z)$, obtained using the numerical solution of the phase plane trajectory (green) with the  $\mathcal{O}(c)$ and $\mathcal{O}(c^2)$ perturbation solutions in yellow and blue, respectively.  Results in (g)--(i) show magnified comparison of the three solutions in the regions highlighted by the dashed boxes in (d)--(f).}
		\label{fig:7}
	\end{figure}
\end{landscape}

Results in Figure \ref{fig:7} for the receding travelling wave are presented in the exact same format as those in Figure \ref{fig:6}.  Here, in Figure \ref{fig:6} we consider $c=-0.5, -0.75$ and $-1.00$ and we see that the $\mathcal{O}(c^2)$ perturbation solution provides a very accurate approximation of both the phase plane trajectory and the shape of the receding travelling wave.

As we pointed out previously, one of the key conceptual limitations of using the Fisher--Stefan model is that, unlike applications in physical and material sciences~\cite{Crank1987,Dalwadi2020,Hill1987,Mitchell2014}, estimates of $\kappa$ are not available.  One way to address this limitation is to use our analysis to provide a relationship between $\kappa$ and $c$, since the wave speed is relatively straightforward to measure~\cite{Maini2004a,Maini2004b,Simpson2007} and could be used to infer an estimate of $\kappa$.  As noted previously, all travelling wave solutions of the Fisher--Stefan model satisfy $\kappa = -c/V(0)$, where $V=V(U)$.  When $|c| \ll 1$ we can estimate $V(0)$ using our perturbation solutions and this provides various relationships between $\kappa$ and $c$ depending on the order of the perturbation solution for $V(0)$,
\begin{align}
\mathcal{O}(1): \kappa & = \dfrac{-c}{V_0(0)}, \label{eq:kappacorder1csmall}\\
\mathcal{O}(c): \kappa & = \dfrac{-c}{V_0(0) + c V_1(0) }, \label{eq:kappacorderccsmall}\\
\mathcal{O}(c^2): \kappa & = \dfrac{-c}{V_0(0) + c V_1(0) + c^2 V_2(0)}. \label{eq:kappacordercsquarecsmall}
\end{align}
Substituting expressions for $V_0(0)$, $V_1(0)$ and $V_2(0)$ and expanding the resulting expressions for $|c| \ll 1$ gives
\begin{align}
\label{eq:kappacO1csmall}
\mathcal{O}(1): \kappa(c) &= \sqrt{3}c +\mathcal{O}(c^2), \\
\mathcal{O}(c): \kappa(c) &= \sqrt{3}c-\dfrac{3}{5}(2-3\sqrt{3})c^2+\mathcal{O}(c^3), \label{eq:kappacOccsmall}\\
\begin{split}
\mathcal{O}(c^2): \kappa(c) &= \sqrt{3}c-\dfrac{3}{5}(2-3\sqrt{3})c^2 \\
& -\dfrac{9\sqrt{3}}{50}\left[10\ln\left(\dfrac{6}{2+\sqrt{3}}\right)+12\sqrt{3}-31\right]c^3 + \mathcal{O}(c^4),
\end{split}
\label{eq:kappacOc2csmall}
\end{align}
which provides a simple way to relate $c$ and $\kappa$ for $|c| \ll 1$.  To explore the accuracy of these approximations we use numerical solutions in the phase plane to estimate $\kappa$ in the interval $-1 < c < 1$ and show the numerically--determined relationship between $c$ and $\kappa$ in Figure \ref{fig:7}.  We also superimpose the various approximations, given by Equations (\ref{eq:kappacO1csmall})--(\ref{eq:kappacOc2csmall}) in Figure \ref{fig:7}, where we see that Equation (\ref{eq:kappacOc2csmall}) is particularly accurate for $|c| \ll 0.5$.

\begin{figure}[H]
	\centering
	\includegraphics[width=1\linewidth]{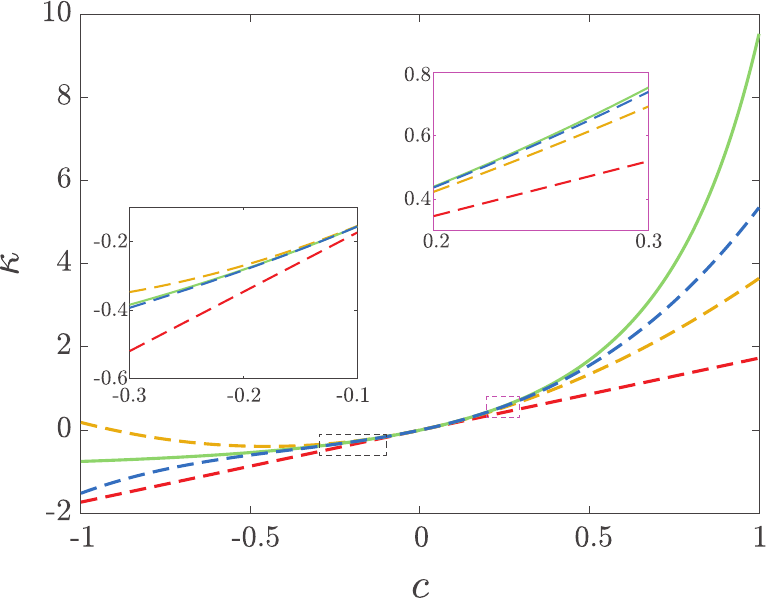}
	\caption{\textbf{Relationship between $c$ and $\kappa$ for $|c| \ll 1$.} The numerical estimate of $\kappa$ as a function of $c$ is given in solid green.  Various perturbation approximations given by Equation (\ref{eq:kappacO1csmall})--(\ref{eq:kappacOc2csmall}) are given in dashed red, dashed yellow and dashed blue, respectively.  The various relationships between $c$ and $\kappa$ are shown in two insets.  The first inset, for $-0.3 < c < 0.1$, is outlined in black.  The second inset, for $0.2 < c < 0.3$, is outlined in pink.}
	\label{fig:8}
\end{figure}

\subsubsection{Perturbation solution for fast receding travelling waves } \label{sec:MatchedAsymptotic}
As noted in Section \ref{sec:NumSims}, preliminary numerical simulations of receding travelling waves in Figure \ref{fig:3}(e)--(h) suggest the formation of a boundary layer as the speed $c$ decreases.  The second order boundary value problem governing the shape of these travelling waves can be written as
\begin{equation} \label{eq:ODEUzepsilon}
\dfrac{1}{c}\frac{\mathrm{d}^2 U}{\mathrm{d} z^2} + \frac{\mathrm{d} U}{\mathrm{d} z} + \dfrac{1}{c}U(1-U) = 0,  \quad  -\infty < z < 0,
\end{equation}
which is singular as $c \to -\infty$.   Therefore, we will construct a matched asymptotic expansion~\cite{Murray1984} by treating $1/c$ as a small parameter.  The boundary conditions  for this problem are $U(0)=0$ and $U(z)=1$ as $z \rightarrow -\infty$.  Setting $1/c =0$ and solving the resulting ODE gives the outer solution,
\begin{equation}
	U(z) = 1,
\end{equation}
which matches the boundary condition as $z \rightarrow -\infty$. To construct the inner solution near $z=0$ we rescale the independent variable $\zeta=zc$. Therefore, in the boundary layer we have
\begin{equation} \label{eq:ODEUxiepsilon}
\frac{\mathrm{d}^2 U}{\mathrm{d} \zeta^2} + \frac{\mathrm{d} U}{\mathrm{d} \zeta} + \dfrac{1}{c^2} U(1-U) = 0,  \quad  -\infty < \zeta < 0.
\end{equation}
Now expanding $U(\zeta)$ in a series we obtain
\begin{equation}
\label{eq:perturbationserieszeta}
U(\zeta) = U_0(\zeta) + \dfrac{1}{c^2} U_1(\zeta) + \dfrac{1}{c^4} U_2(\zeta) + \mathcal{O}\left(\dfrac{1}{c^6}\right),
\end{equation}
which we substitute into Equation (\ref{eq:ODEUxiepsilon}) to give a family of boundary value problems,
\begin{align} \label{eq:ODEU0}
&\frac{\mathrm{d}^2 U_0}{\mathrm{d} \zeta^2} + \frac{\mathrm{d} U_0}{\mathrm{d} \zeta} = 0, & U_0(0) = 0, \ U_0 \rightarrow 1  \ \text{as} \ \zeta \rightarrow -\infty,\\
\label{eq:ODEU1}
&\frac{\mathrm{d}^2 U_1}{\mathrm{d} \zeta^2} + \frac{\mathrm{d} U_1}{\mathrm{d} \zeta} + U_0(1-U_0) = 0,  & U_1(0) = 0, \ U_1 \rightarrow 0   \ \text{as} \ \zeta \rightarrow -\infty,\\
\label{eq:ODEU2}
&\frac{\mathrm{d}^2 U_2}{\mathrm{d} \zeta^2} + \frac{\mathrm{d} U_2}{\mathrm{d} \zeta} + U_1(1-2U_0) = 0,  & U_2(0) = 0, \ U_2 \rightarrow 0  \ \text{as} \ \zeta \rightarrow -\infty.
\end{align}
The solution of these boundary value problems are
\begin{align}
&U_0(\zeta) = (1-e^{-\zeta}) \label{eq:perturbationU0zeta}, \\
&U_1(\zeta) = \left(-\frac{1}{2}+\zeta\right) e^{-\zeta} + \frac{1}{2} e^{-2\zeta},  \label{eq:perturbationU1zeta}\\
&U_2(\zeta) = \frac{e^{-\zeta}}{12}\left[11-e^{-\zeta}\left(9+2e^{-\zeta}\right)\right] -\zeta e^{-\zeta} \left(e^{-\zeta}+\frac{1}{2}\zeta+\frac{1}{2}\right);
\label{eq:perturbationU2zeta}
\end{align}
Maple code to generate these solutions is available on \href{https://github.com/maudelhachem/El-Hachem2020b}{GitHub}.  Combining the inner and outer solution leads to $U(z) = U_0(z) + c^{-2} U_1(z) + c^{-4} U_2(z) + \mathcal{O}(c^{-6})$, where $U_0(z)$, $U_1(z)$, $U_2(z)$ correspond to Equations (\ref{eq:perturbationU0zeta})--(\ref{eq:perturbationU2zeta}), respectively, written in terms of the original variable $z=\zeta/c$. By truncating this series at different orders we are able to compare $\mathcal{O}(1)$, $\mathcal{O}(c^{-2})$ and $\mathcal{O}(c^{-4})$ perturbation solutions.

\begin{figure}[H]
	\centering
	\includegraphics[width=1\linewidth]{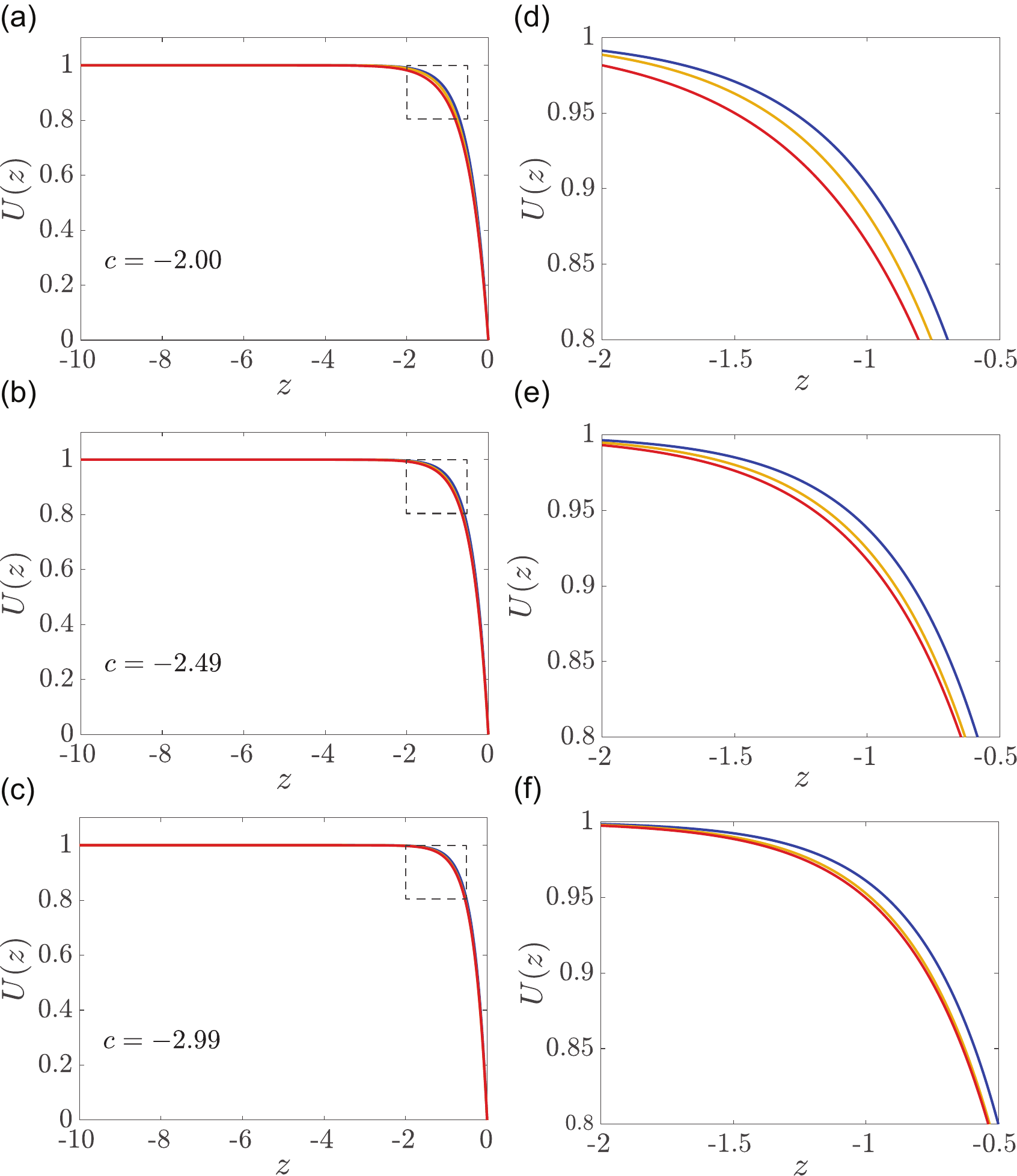}
	\caption{\textbf{Perturbation solutions for slow receding travelling waves.} (a)--(c) show plots of the shape of the travelling waves for $c = -2.00, -2.49$ and $-2.99$, respectively.  Late time numerical solutions of Equations (\ref{eq:NondimPDE})--(\ref{eq:IC}) are shown in blue, and the $\mathcal{O}(1)$ and $\mathcal{O}(c^{-2})$ perturbation solutions are plotted in red and yellow, respectively.  (d)--(f) show the magnified regions highlighted by the dashed boxes in (a)--(c), respectively.}
	\label{fig:9}
\end{figure}

Results in Figure \ref{fig:9} compare the numerical solutions of Equations (\ref{eq:NondimPDE})--(\ref{eq:IC}) with various perturbation solutions for fast receding travelling waves.  Results in Figure \ref{fig:9}(a)--(c) show late--time numerical solutions of the PDE model in blue with $c=-2.00, -2.49$ and $-2.99$, respectively.  In each subfigure, the $\mathcal{O}(1)$ and $\mathcal{O}(c^{-2})$ perturbation solutions are plotted, in red and yellow, respectively.  For these results we have not plotted the $\mathcal{O}(c^{-4})$ perturbation solution in order to keep Figure \ref{fig:9} easy to interpret.  As expected we see that the match between the numerical and perturbation solutions improves as $c$ decreases, and we see that the $\mathcal{O}(c^{-2})$ perturbation solutions are more accurate than the $\mathcal{O}(1)$ perturbation solutions.  Results in Figure \ref{fig:9}(d)--(f) show a magnified comparison of the three solutions and the regions shown are highlighted in the dashed box in Figure \ref{fig:9}(a)--(c).

\begin{figure}[H]
	\centering
	\includegraphics[width=1\linewidth]{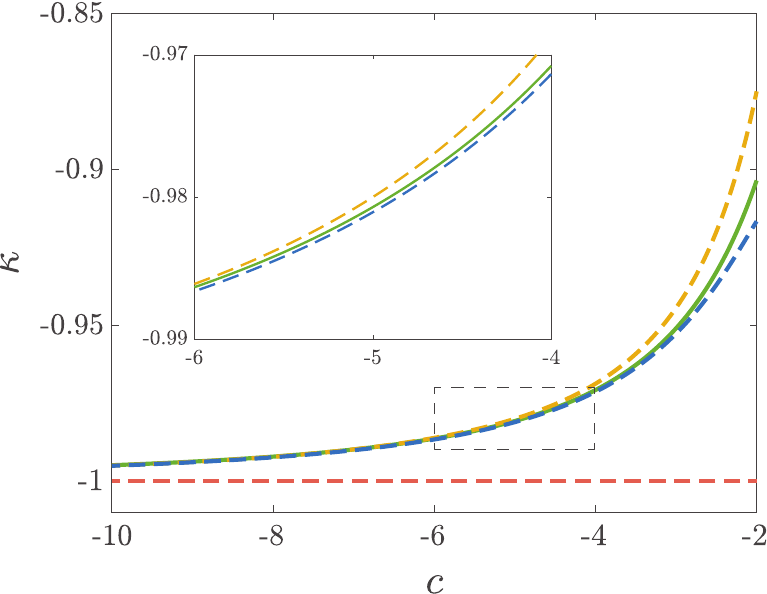}
	\caption{\textbf{Relationship between $c$ and $\kappa$ near $c \to -\infty$.} The numerical estimate of $\kappa$ as a function of $c$ is given in solid green.  Various perturbation approximations given by Equation (\ref{eq:kappacO1cbig})--(\ref{eq:kappacOcm4cbig}) are given in dashed red, dashed yellow and dashed blue, respectively.  Various relationships between $c$ and $\kappa$ are shown in an inset, for $-6 < c < -4$.}
	\label{fig:10}
\end{figure}

For all travelling wave solutions we have $\kappa = -c/V(0)$. As $c \to -\infty$ we can estimate $V(0)$ using our perturbation solutions to provide insight into the relationship between $\kappa$ and $c$.  We achieve this by evaluating the following expressions,
 \begin{align}
\mathcal{O}(1): \kappa &= \dfrac{-c}{\dfrac{\mathrm{d} U_0(0)}{\mathrm{d} z}} , \label{eq:kappacorder1cbig}\\
\mathcal{O}\left(\dfrac{1}{c^2}\right): \kappa &= \dfrac{-c}{\dfrac{\mathrm{d} U_0(0)}{\mathrm{d} z} + \dfrac{1}{c^2}\dfrac{\mathrm{d}U_1(0)}{\mathrm{d} z}}, \label{eq:kappacordercm2cbig}\\
\mathcal{O}\left(\dfrac{1}{c^4}\right): \kappa &= \dfrac{-c}{\dfrac{\mathrm{d} U_0(0)}{\mathrm{d} z} + \dfrac{1}{c^2}\dfrac{\mathrm{d}U_1(0)}{\mathrm{d} z} + \dfrac{1}{c^4}\dfrac{\mathrm{d}U_2(0)}{\mathrm{d} z}},\label{eq:kappacordercm4cbig}
\end{align}
where we must differentiate our expressions for $U_0(z)$, $U_1(z)$ and $U_2(z)$ with respect to $z$. Substituting our perturbation solutions into Equations (\ref{eq:kappacorder1cbig})--(\ref{eq:kappacordercm4cbig}) and then expanding the resulting terms as $c \to -\infty$ gives
\begin{align}
\label{eq:kappacO1cbig}
\mathcal{O}(1): \kappa(c) &= -1 + \mathcal{O}\left(\dfrac{1}{c^2}\right),\\
\mathcal{O}\left(\dfrac{1}{c^2}\right): \kappa(c) &= -1+\frac{1}{2c^2}+\mathcal{O}\left(\dfrac{1}{c^4}\right), \label{eq:kappacOcm2cbig}\\
\mathcal{O}\left(\dfrac{1}{c^4}\right): \kappa(c) &= -1+\frac{1}{2c^2}-\dfrac{2}{3c^4}+\mathcal{O}\left(\dfrac{1}{c^6}\right), \label{eq:kappacOcm4cbig}
\end{align}
which provides us with a simple way to relate $\kappa$ and $c$ as $c \to -\infty$.  To explore the accuracy of these approximations we use numerical solutions in the phase plane to estimate $\kappa$ in the interval $-10 < c < -2$ and show the numerically--determined relationship between $c$ and $\kappa$ in Figure \ref{fig:10}.  We also superimpose the various approximations, given by Equations (\ref{eq:kappacO1cbig})--(\ref{eq:kappacOcm4cbig}) in Figure \ref{fig:10}, where we see that $\kappa \to -1^+$ as $c \to -\infty$, and that Equation (\ref{eq:kappacOcm4cbig}) gives an excellent approximation of $\kappa$ for $c < -2$.

In summary, in Sections \ref{sec:Exactsolution}--\ref{sec:MatchedAsymptotic} we provide analysis for the case of $c=0$, $|c| \ll 1$ (slowly invading or slowly receding) and $-c  \gg 1$ (fast receding), respectively.  It is also possible to analyse the special case where $c = - \sqrt{5}/6$, where the solution can be written in terms of Weierstrass elliptic functions~\cite{McCue2020}.

\section{Conclusion and Outlook} \label{sec:Conclusion}
In this work we discuss approaches for modelling biological invasion and recession.  The most commonly--used model to mimic biological invasion is the Fisher--KPP model~\cite{Edelstein2005,Murray2002}, and generalisations of the Fisher--KPP model, such as the Porous--Fisher model~\cite{Murray2002,Witelski1995}.  While these single--species PDE models have been used to simulate biological invasion in various contexts, they cannot be used to simulate biological recession.  As an alternative, we explore the Fisher--Stefan model~\cite{Du2010,Elhachem2019}, which is a different generalisation of the Fisher--KPP model obtained by reformulating the classical model as a moving boundary problem.

There are both advantages and disadvantages of reformulating the Fisher--KPP model as a moving boundary problem.  One advantage of using the Fisher--Stefan model is that it involves a well--defined sharp front and it has the ability to model both biological invasion and recession.  These advantages are both attractive because experimental observations of biological invasion typically report well--defined sharp fronts~\cite{Maini2004a,Maini2004b} and it is well--known that motile and proliferative populations can both invade and recede.  The Fisher--KPP model cannot describe either of these observed features.  A disadvantage of using the Fisher--Stefan model is the need to specify the constant, $\kappa$.  While estimates of these kinds of parameters are well--known in the heat and mass transfer literature for modelling physical processes~\cite{Crank1987,Dalwadi2020,Hill1987,Mitchell2014}, there are no such estimates for these parameters in a biological or ecological context that we are aware of.  Part of the motivation for the analysis in this work is to provide numerical and approximate analytical insight into the relationship between $\kappa$ and $c$.  We are motivated to do this because measurements of $c$ are often reported~\cite{Maini2004a,Maini2004b,Simpson2007} and so understanding how to interpret an estimate of $c$ in terms of $\kappa$ is of interest. \cbl In summary, we show that slowly invading or receding travelling wave solutions of the Fisher-Stefan model move with speed $c \sim \kappa / \sqrt{3}$ as $\kappa \to 0$, whereas rapidly receding travelling wave solutions of the Fisher-Stefan model move with speed $c \sim 2^{-1} (\kappa+1)^{-1/2}$ as $\kappa \to -1^+$. \cb

In this work we compare the Fisher--KPP model and the Fisher--Stefan model and it is interesting to consider how these models can be used to interpret experimental observations.  As discussed, experimental estimates of $c$ are the most straightforward measurement to obtain in cell biology experiments.  For example, Maini et al.~\cite{Maini2004a} use a scratch assay to obtain an estimate of $\hat{c}$, whereas Simpson et al.~\cite{Simpson2007} report estimates of $\hat{c}$ using observations of cell invasion within intact embryonic tissues.  With these measurements of $\hat{c}$, it is possible to estimate the product of the diffusivity and the proliferation rate since $\hat{c} = 2\sqrt{\hat{\lambda}\hat{D}}$ for the Fisher--KPP model.  A standard practice is to infer $\hat{\lambda}$ by assuming that a typical doubling time is, say, $24$ h, giving  $\hat{\lambda} = \textrm{ln}(2)/24$ /h.  These two pieces of information can be used to estimate $\hat{D}$ by assuming that travelling wave solutions of the Fisher--KPP model are relevant and  $\hat{c} = 2\sqrt{\hat{\lambda}\hat{D}}$.  This approach was followed by Maini et al.~\cite{Maini2004a,Maini2004b} and Simpson et al.~\cite{Simpson2007}.  Unfortunately this simple approach does not provide any certainty that the Fisher--KPP model is actually valid.  Indeed, with more experimental effort it is possible to carefully analyse a cell proliferation assay to provide a separate estimate of $\hat{\lambda}$~\cite{Browning2017}, and to either track individual cells~\cite{Cai2007} or to chemically--inhibit proliferation~\cite{Simpson2013} to obtain an independent estimate of $\hat{D}$.  If these more careful experiments are performed, it is then possible to examine if the relationship $\hat{c} = 2\sqrt{\hat{\lambda}\hat{D}}$ is indeed true.  If this classical relationship does not hold and $\hat{c} < 2\sqrt{\hat{\lambda}\hat{D}}$,  the Fisher--Stefan model provides a better explanation of the data since it is always possible to choose a value of $\hat{\kappa}$ to match independent estimates of $\hat{D}$, $\hat{\lambda}$ and $\hat{c}$.

In conclusion we would like to mention that all of the models discussed in this work make the very simple but extremely common assumption that the proliferation of individuals is given by a logistic source term.  This assumption is widely invoked in many single species models of invasion, including the Fisher--KPP model~\cite{Maini2004a,Maini2004b,Simpson2007}, the Porous--Fisher model~\cite{Buenzli2020,Sherratt1990,Witelski1995} and the Fisher--Stefan model~\cite{Du2010,Elhachem2019}, as well as many more complicated multiple species analogues of these models~\cite{Chaplain2020,Painter2003,Painter2015}.  We acknowledge that there are other classes of models where different source terms are used, such as the bistable equation and various models that describe Allee effects~\cite{Courchamp2008,Fadai2020b,Fife1979,Johnston2017,Lewis1993,Taylor2005}.  These models are similar to the classical Fisher--KPP model except that the quadratic source term is generalised to a cubic source term, and it is well--known that such single species models can be used to simulate both biological and invasion and retreat by changing the shape of the cubic source term.  In this work we have deliberately not focused on Allee--type models so that we do not conflate models of Allee effects with the Fisher--Stefan model.  Of course, it would be very interesting to consider an extension of the Fisher--Stefan model with a more general source term~\cite{Browning2017,Tsoularis2002}, such as an Allee effect.  We anticipate many of the numerical, phase plane and perturbation tools developed in this work would also play a role in the analysis of a Fisher--Stefan--type model with a generalised source term.  We leave this extension for future consideration.

\newpage

\section*{Appendix A: Numerical methods } \label{sec:Numericalmethods}

\subsection{Partial differential equation}
\label{sec:PDE}

To obtain numerical solutions of the Fisher--Stefan equation
\begin{equation}\label{eq:FisherKPPmovbound}
\frac{\partial u}{\partial t} =\frac{\partial^2 u}{\partial x^2} +  u(1-u),
\end{equation}
for $0 < x < L(t)$ and $t > 0$, we first use a boundary fixing transformation $\xi = x / L(t)$ so that we have
\begin{align}\label{eq:FisherKPPmovboundxi}
\frac{\partial u}{\partial t} = \frac{1}{L^2(t)} \frac{\partial^2 u}{\partial \xi^2}+\frac{\xi}{L(t)} \frac{\text{d}L(t)}{\text{d} t} \frac{\partial u}{\partial \xi} + u(1-u),
\end{align}
on the fixed domain, $0 < \xi < 1$, for $t > 0$.  Here $L(t)$ is the length of the domain that we will discuss later.  To close the problem we  also transform the boundary conditions giving
\begin{align}
&\dfrac{\partial u}{\partial \xi} = 0 \quad \textrm{at} \quad \xi=0,  \label{eq:FS_BC1} \\
&u = 0 \quad \textrm{at} \quad \xi=1. \label{eq:FS_BC2}
\end{align}

We spatially discretise Equations (\ref{eq:FisherKPPmovboundxi})--(\ref{eq:FS_BC2}) with a uniform finite difference mesh, with spacing $\Delta \xi$, approximating the spatial derivatives using a central finite difference approximation, giving
\begin{align}\label{eq:FDDinternalxi}
\dfrac{u_{i}^{j+1} - u_{i}^{j} }{\Delta t} &=  \dfrac{1}{(L^{j})^2} \left( \frac{u_{i-1}^{j+1} - 2 u_{i}^{j+1} + u_{i+1}^{j+1}}{\Delta \xi^2} \right)\notag \\
&+ \dfrac{\xi}{L^{j}} \left(\dfrac{L^{j+1} - L^{j} }{\Delta t}\right) \left( \frac{u_{i+1}^{j+1} - u_{i-1}^{j+1}}{2 \Delta \xi} \right) +  u_{i}^{j+1}(1 - u_{i}^{j+1}) ,
\end{align}
for $i = 2, \ldots, m-1$, where $m = 1/\Delta \xi + 1$ is the total number of spatial nodes on the finite difference mesh, and the index $j$ represents the time index so that $u_{i}^{j} \approx u(\xi,t)$, where $\xi~=~(i~-~1)~\Delta~\xi$ and $t=j\Delta t$.

Discretising Equations (\ref{eq:FS_BC1})--(\ref{eq:FS_BC2}) leads to
\begin{align}
&u_{2}^{j+1}-u_{1}^{j+1} = 0,  \label{eq:FS_BC1a} \\
&u_{m}^{j+1} = 0. \label{eq:FS_BC2a}
\end{align}

To advance the discrete system from time $t$ to $t + \Delta t$ we solve the system of nonlinear algebraic equations, Equations (\ref{eq:FDDinternalxi})-(\ref{eq:FS_BC2a}), using Newton-Raphson iteration. During each iteration of the Newton--Raphson algorithm we estimate the position of the moving boundary using the discretised Stefan condition,
\begin{equation}
L^{j+1} = L^{j} - \dfrac{\Delta t  \kappa}{L^j} \left(\dfrac{u_{m}^{j+1}-u_{m-1}^{j+1}}{\Delta \xi}\right).
\label{eq:Lupdatediscretise}
\end{equation}
Within each time step the Newton--Raphson iterations continue until the maximum change in the dependent variables is less than the tolerance  $\epsilon$.  All results in this work are obtained by setting $\epsilon = 1 \times 10^{-8}$, $\Delta \xi = 1 \times 10^{-6}$ and $\Delta t = 1 \times 10^{-2}$, and we find that these values are sufficient to produce grid--independent results.  However, we recommend that care be taken when using the algorithms on \href{https://github.com/maudelhachem/El-Hachem2020b}{GitHub} when considering larger values of $\kappa$, which can require a much denser mesh to give grid--independent results.

We use the time--dependent solutions to provide an estimate of the travelling wave speed $c^*$. The estimated wave speed is computed using the discretised position of the moving boundary such as $c^* = (L^{j+1} - L^{j})/\Delta t$.

\subsection{Phase plane}
\label{PhasePlane}
To construct the phase planes we solve Equations (\ref{eq:ODEdU})--(\ref{eq:ODEdV}) numerically using Heun's method with a constant step size $\textrm{d}z$. In most cases we are interested in examining trajectories that either enter or leave the saddle $(1,0)$ along the stable or unstable manifold, respectively.  Therefore, it is important that the initial condition we chose when solving Equations (\ref{eq:ODEdU})--(\ref{eq:ODEdV}) are on the appropriate stable or unstable manifold and sufficiently close to $(1,0)$.  To choose this point we use the MATLAB \textit{eig} function~\cite{eig} to calculate the eigenvalues and eigenvectors for the particular choice of $c$ of interest. The flow of the dynamical system are plotted on the phase planes using the MATLAB \textit{quiver} function~\cite{quiver}.

\newpage
 \section*{Appendix B: Time--dependent PDE solutions with different initial conditions} \label{sec:Additional}
\begin{figure}[H]
	\centering
	\includegraphics[width=0.7\linewidth]{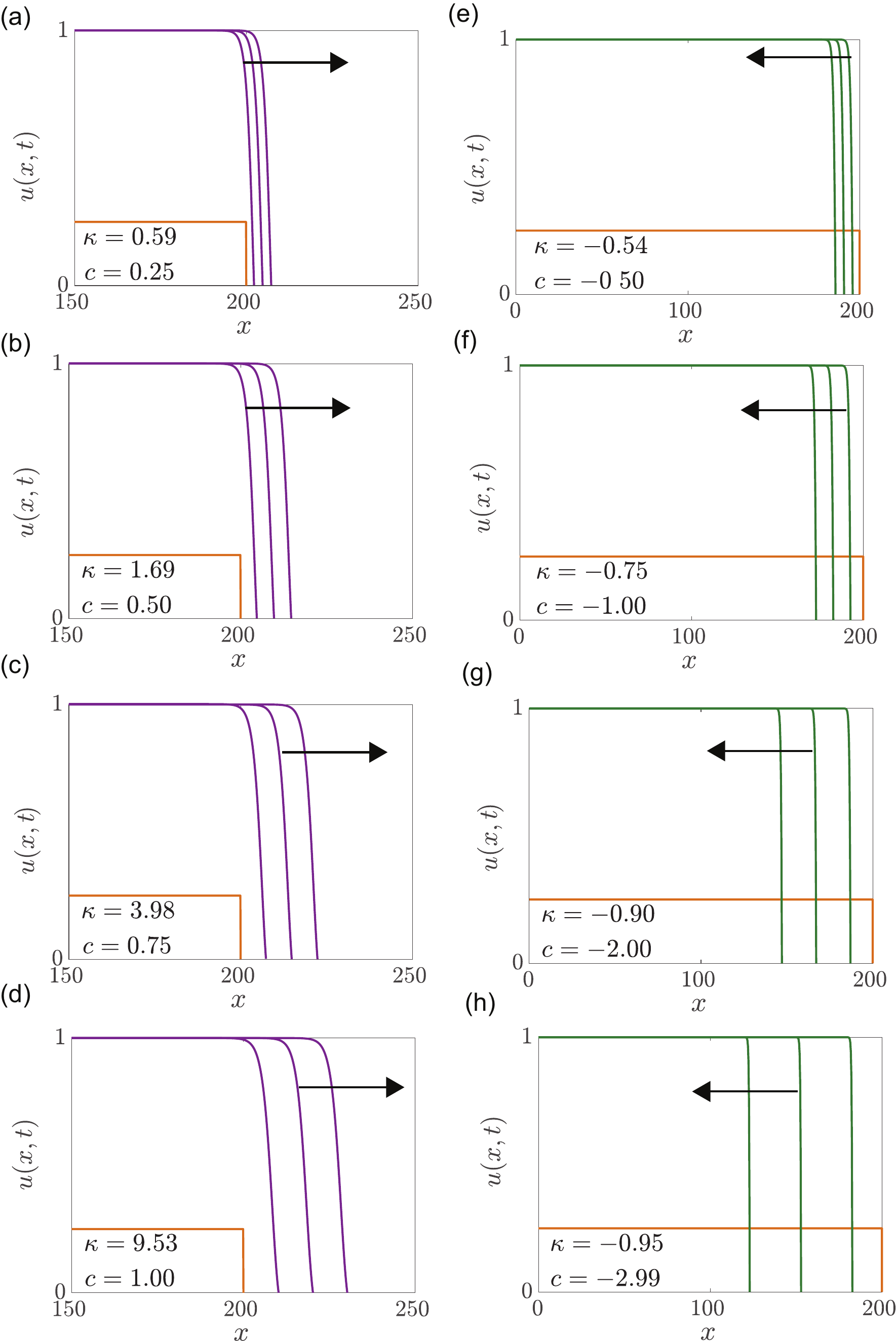}
	\caption{\textbf{Invading and receding travelling wave solutions of the Fisher--Stefan model.} Numerical solutions of Equations (\ref{eq:NondimPDE})--(\ref{eq:IC}) are given at $t = 0,10,20$ and $30$. The initial condition is given by Equation (\ref{eq:IC}) with $\alpha=0.25$ and $L(0)=200$.  Results in (a)--(d) lead to invading travelling waves with $c=0.25,0.50,0.75$ and $1.00$, respectively.  These travelling waves are obtained by choosing $\kappa = 0.5859, 1.6879, 3.9823$ and $9.5315$, respectively.  Results in (e)--(h) lead to receding travelling waves with $c=-0.50,-1.00,-2.00$ and $-2.99$, respectively. These receding travelling waves are obtained by choosing $\kappa = -0.5387, -0.7529, -0.9036$ and $-0.9510$, respectively.  Our estimates of $c$ correspond are obtained at late time, here $t=30$.  Note that estimates of $\kappa$ are reported in the caption to four decimal places, whereas the estimates given in the subfigures are reported to two decimal places to keep the figure neat.}
	\label{fig:B1}
\end{figure}

\begin{figure}[H]
	\centering
	\includegraphics[width=0.7\linewidth]{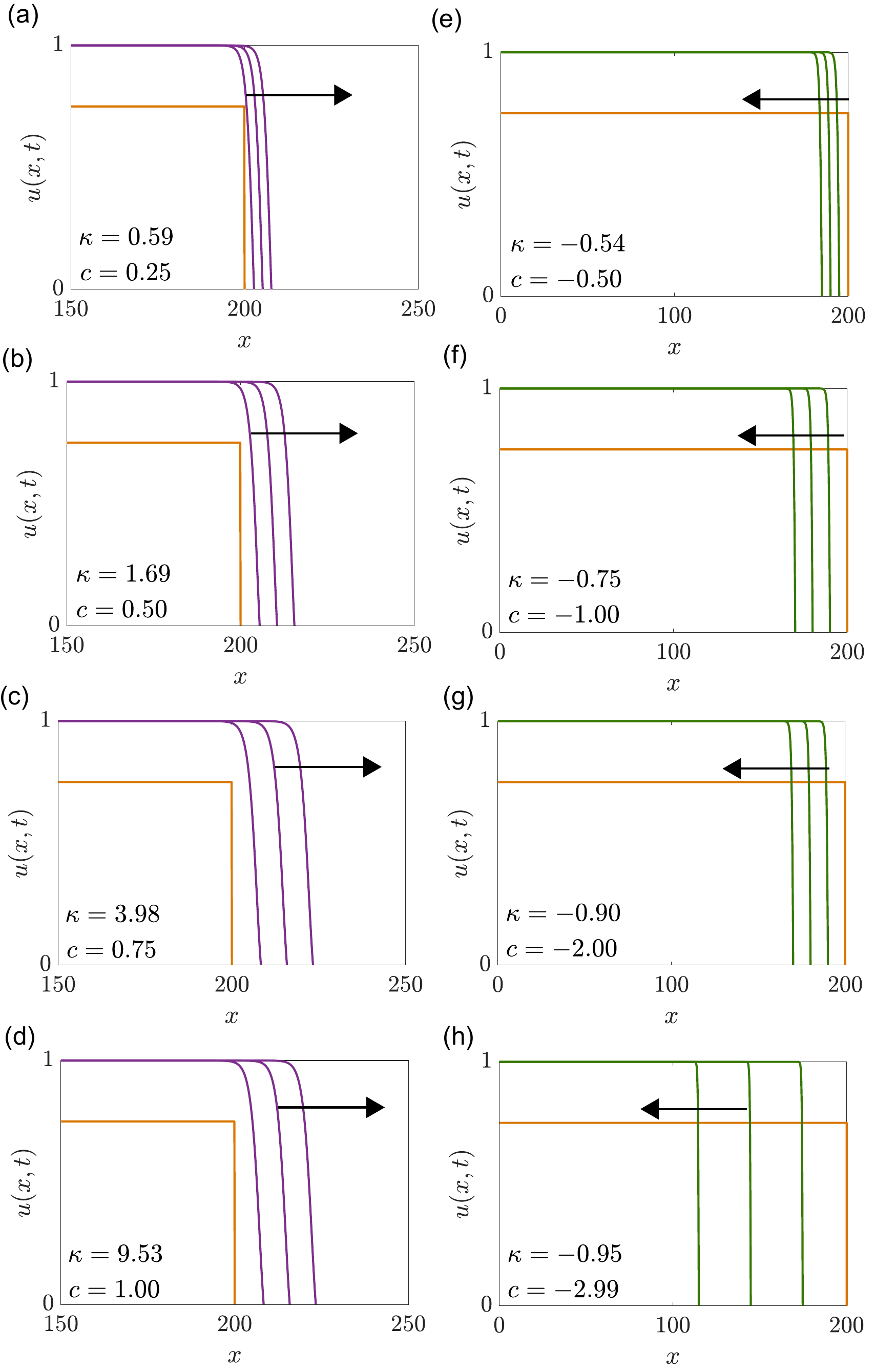}
	\caption{\textbf{Invading and receding travelling wave solutions of the Fisher--Stefan model.}  Numerical solutions of Equations (\ref{eq:NondimPDE})--(\ref{eq:IC}) are given at $t = 0,10,20$ and $30$. The initial condition is given by Equation (\ref{eq:IC}) with $\alpha=0.75$ and $L(0)=200$.  Results in (a)--(d) lead to invading travelling waves with $c=0.25,0.50,0.75$ and $1.00$, respectively.  These travelling waves are obtained by choosing $\kappa = 0.5859, 1.6879, 3.9823$ and $9.5315$, respectively.  Results in (e)--(h) lead to receding travelling waves with $c=-0.50,-1.00,-2.00$ and $-2.99$, respectively. These receding travelling waves are obtained by choosing $\kappa = -0.5387, -0.7529, -0.9036$ and $-0.9510$, respectively.  Our estimates of $c$ correspond are obtained at late time, here $t=30$.  Note that estimates of $\kappa$ are reported in the caption to four decimal places, whereas the estimates given in the subfigures are reported to two decimal places to keep the figure neat.}
	\label{fig:B2}
\end{figure}

\begin{figure}[H]
	\centering
	\includegraphics[width=0.7\linewidth]{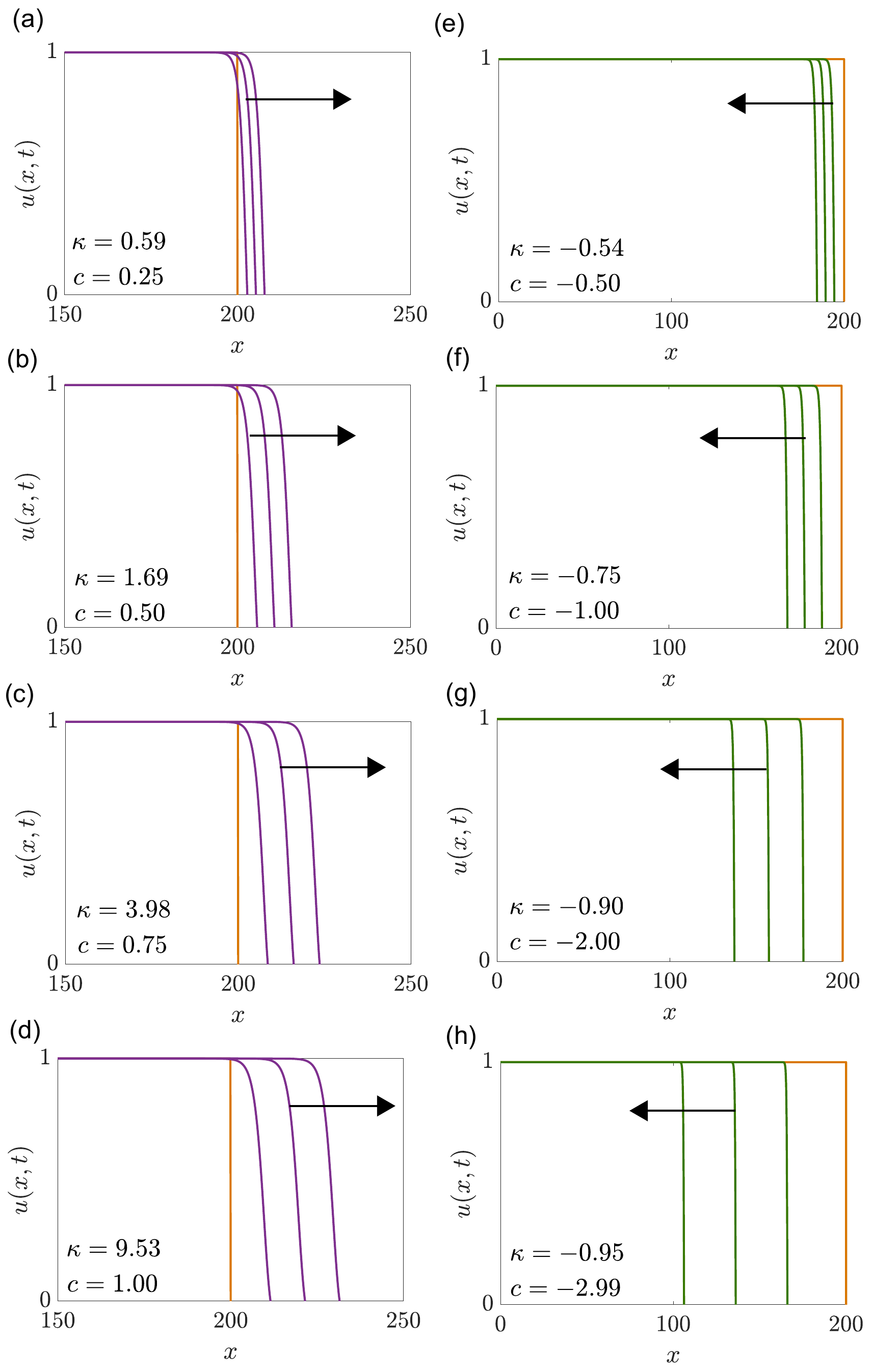}
	\caption{\textbf{Invading and receding travelling wave solutions of the Fisher--Stefan model.}  Numerical solutions of Equations (\ref{eq:NondimPDE})--(\ref{eq:IC}) are given at $t = 0,10,20$ and $30$. The initial condition is given by Equation (\ref{eq:IC}) with $\alpha=1.00$ and $L(0)=200$.  Results in (a)--(d) lead to invading travelling waves with $c=0.25,0.50,0.75$ and $1.00$, respectively.  These travelling waves are obtained by choosing $\kappa = 0.5859, 1.6879, 3.9823$ and $9.5315$, respectively.  Results in (e)--(h) lead to receding travelling waves with $c=-0.50,-1.00,-2.00$ and $-2.99$, respectively. These receding travelling waves are obtained by choosing $\kappa = -0.5387, -0.7529, -0.9036$ and $-0.9510$, respectively.  Our estimates of $c$ correspond are obtained at late time, here $t=30$.  Note that estimates of $\kappa$ are reported in the caption to four decimal places, whereas the estimates given in the subfigures are reported to two decimal places to keep the figure neat.}
	\label{fig:B3}
\end{figure}

Results in Figure \ref{fig:3} show a family of time--dependent solutions of the Fisher-Stefan model that lead to both invading and receding travelling waves for different choices of $\kappa$, but the same choice of initial condition, Equation (\ref{eq:IC}) with $\alpha = 0.5$.  Here, in Figures \ref{fig:B1}--\ref{fig:B3} we present analogous results except we change the initial condition by choosing $\alpha = 0.25, 0.75$ and 1.00, respectively.   Comparing the shape of the long-time travelling wave solutions in Figure \ref{fig:3} with those here in Figures \ref{fig:B1}--\ref{fig:B3} confirms that the eventual travelling wave solutions are independent of the initial condition.  Here, the time-dependent solution at $t=30$ is sufficient to see this.  For example, in Figure \ref{fig:3}(g) with $\kappa = -0.9$, we eventually see that a receding travelling wave with $c=-2.00$ forms by $t=30$.  Results in Figure \ref{fig:B1}(g), Figure \ref{fig:B2}(g) and Figure \ref{fig:B3}(g) confirm that we obtain the same travelling wave, with the same long time wave speed, regardless of the initial condition.  Of course, should the reader wish to experiment with other choices of initial condition, or if they wish to explore the time--dependent solutions in Figure \ref{fig:3} or Figures \ref{fig:B1}--\ref{fig:B2} for a longer duration of time, say $t=40$, they may do so by downloading and running the MATLAB code provided on \href{https://github.com/maudelhachem/El-Hachem2020b}{GitHub}. \cb

\paragraph{Acknowledgements:} We thank Stuart Johnston and Sean McElwain for helpful suggestions and feedback. This work is supported by the Australian Research Council (DP200100177).

\end{document}